\documentclass[a4paper,fleqn]{cas-sc}

\usepackage[authoryear]{natbib}
\usepackage{lineno}

\usepackage{xcolor}
\usepackage[acronym]{glossaries}
\usepackage{listings}
\usepackage{subfig}
\usepackage[toc,page]{appendix}
\usepackage{array}

\begin{document}
\let\WriteBookmarks\relax
\def\floatpagepagefraction{1}
\def\textpagefraction{.001}

\shorttitle{Diagnostic vs dynamic IB in a global ocean model}    

\shortauthors{{NMK, KM, S-JL, {\O}B}}  

%\title [mode = title]{Evaluating two methods for including the inverse barometer effect in a global ocean model}   
\title [mode = title]{Diagnostic vs dynamic representation of the inverse barometer effect in a global ocean model and its potential for probabilistic storm surge forecasting}   
%\tnotemark[<tnote number>] 

%\tnotetext[<tnote number>]{<tnote text>} 


\author[1]{Nils Melsom Kristensen}[orcid=0000-0002-2494-6509]
\cormark[1]
\cortext[1]{Corresponding author}
\credit{Software, Formal analysis, Validation, Visualization, Writing – original draft, Writing – review \& editing}
\ead{nilsmk@met.no}

\author[3]{Kristian Mogensen}[orcid=0000-0002-8773-3017]
\credit{Conceptualization, Methodology, Investigation, Data curation, Writing – review \& editing}

\author[3]{Sarah-Jane Lock}[orcid=0000-0003-0936-8170]
\credit{Conceptualization, Methodology, Investigation, Data curation, Writing – review \& editing}

\author[1,2]{{\O}yvind Breivik}[orcid=0000-0002-2900-8458]
\credit{Conceptualization, Funding acquisition, Project administration, Writing – original draft, Writing – review \& editing}
\affiliation[1]{organization={Norwegian Meteorological Institute},
            addressline={P.O. Box 43 Blindern}, 
            city={Oslo},
            postcode={NO-0313}, 
            country={Norway}}

\affiliation[2]{organization={University of Bergen},
            city={Bergen},
            country={Norway}}

\affiliation[3]{organization={European Centre for Medium-Range Weather Forecasts},
            city={Reading},
            country={UK}}


%\linenumbers
% Here goes the abstract, max 250 words
\begin{abstract}
The global ocean model NEMO is run in a series of stand-alone configurations (2015--2022) to investigate the potential for improving global medium-range storm surge forecasts by including the inverse barometer effect. The analysis focus on the residual water level, i.e. the water level variations not due to tides. Here, we compare a control experiment, where the inverse barometer effect was not included, against a run dynamically forced with mean sea level pressure. In the control experiment, the inverse barometer effect was then calculated diagnostically and added to the ocean model sea surface elevation, resulting in a total of three experiments to investigate. We compare against the global GESLA3 water level data set and find that the inclusion of the inverse barometer effect reduces the root-mean-square error by $\sim 1$~cm on average.
When we mask out all data where the observed residual water level is less than $\pm1$ or $\pm2$ standard deviations, including the inverse barometer effect reduces the RMS error by $4-5$ cm. 
While both methods reduce water level errors, there are regional differences in their performance. The run with dynamical pressure forcing is seen to perform slightly better than diagnostically adding the inverse barometer effect in enclosed basins such as the Baltic Sea. Finally, an ensemble forecast experiment with the Integrated Forecast System of the European Centre for Medium-range Weather Forecasts demonstrates that when the diagnostic inverse barometer effect is included for a severe storm surge event in the North Sea (Storm Xaver, December 2013), the ensemble spread of water level provides a stronger and earlier indication of the observed maximum surge level than the when the effect is excluded.
\end{abstract}

% Keywords, max 7
\begin{keywords}
Water level  \sep Residual water level \sep Storm surge \sep Inverse barometer effect \sep Ensemble prediction systems
\end{keywords}

\maketitle

% Main text
\section{Introduction}\label{sec:intro}
The mean sea level varies on time scales ranging from hours (tides and synoptic weather) to decades (climate change and large-scale climate fluctuations) \citep{gregory2019}. In addition to the steric changes (expansion and contraction due to changes in the heat content and the salinity of the ocean interior), changes to the meltwater input affect the long-term trends in still water level. On much shorter time scales, we have the synoptic fluctuations caused by variations in the sea level pressure of the overlying atmosphere, known as the inverse barometer (IB) effect \citep{wunsch1997}. While typically isostatic, dynamic behavior arises at specific wave frequencies. \cite{ponte1991} found that global departures from the pure IB response are usually small ($1-3~\mathrm{cm}$), but increase to $5\%-20\%$ for periods between 2 and 7 days. Furthermore, pressure at high latitudes can drive sea level changes at lower latitudes, a nonlocal effect that local IB (diagnostic) corrections miss. These studies demonstrate that dynamic forcing is essential to accurately model high-frequency water level changes.

For the IB effect, a decrease in the surface pressure of 1 hPa leads to a local increase in water level of approximately 1 cm is a practical rule of thumb allowing quick estimates of the impact of atmospheric pressure on the water level \citep{piecuch2015}. In addition, frictional effects will lead to water level variations that have a much wider geographical extent. The simplest model on a rotating sphere is the Ekman transport \citep{gil82}[pp 319--322], where a balance is reached between the frictional force of the wind on the sea surface and the Coriolis effect. When the water is pushed against a coastline, or drawn away from the coastline, the effect typically sets up Kelvin waves trapped by the coast. Such water level variations can be substantial, and are typically among the most devastating effects of tropical cyclones. The extratropics also experience large fluctuations due to Ekman effects, dramatically demonstrated by, among other cases, the devastating storm surge of 1953 in the North Sea \citep{wolf05}. We note that Ekman transport and the IB effect does not always act simply as additive components, but can sometimes be dynamically coupled, as shown by \cite{gomis2008}. They show how, in certain regions, wind forcing can sometimes counteract and override the IB response.

Storm surge modeling is a routine activity of national hydro-meteorological services (see e.g. \cite{kristensen22}) and can be carried out using modeling systems of varying degrees of complexity, ranging from two-dimensional homogeneous equations of the vertically averaged (barotropic) current and its associated water level to more complex fully baroclinic three dimensional ocean models. Such models are run on regional scales and in some cases globally \citep{cirano2025}. The models can include the major constituents of the astronomical tide, or these can be added diagnostically. Regional forecast systems typically rely on boundary conditions from basin-scale or global model systems. Their horizontal resolution is high enough (around 4 km is a typical number) that local topographic effects can be accounted for. Since variations in water level are mostly a concern along coastlines (but not only, fixed platforms must account for changes in the water level, and floating structures may have to account for tension on moorings), such forecast systems are expected to yield reasonably accurate water level predictions in semi-enclosed waters. Furthermore, the uncertainty associated with the passage of synoptic weather systems translates into uncertainty of the water level they cause. Ensemble prediction is thus a natural approach for water level forecasting.

In addition to dedicated storm surge models, any ocean model with a free-surface, which includes most global and regional ocean models, is capable of predicting water levels. However, most global, and some of the larger regional, models do not include the water level contribution from the IB effect (static IB assumption). The present work seeks to explore whether this shortcoming can be overcome by adding the IB effect as a post process, or if it must be included at run-time. 

The limitations of the static IB assumption, specifically its failure at periods shorter than 3 days and in high-latitude regions, have been well-documented. \cite{carrere2003} demonstrated that a global barotropic model (MOG2D-G) using a finite element unstructured mesh (ranging from 400 km in the deep ocean to 20 km in coastal areas) could reduce sea level variance by more than 50\% at tide gauge locations compared to the classical IB correction. A critical finding of their work was that while dynamic pressure forcing alone offered modest improvements, the inclusion of wind stress was responsible for the majority of the variance reduction.

We illustrate the potential for local storm surge modelling, by exploring the impact in the global Integrated Forecast System (IFS) ensemble of the European Centre for Medium-range Weather Forecasts (ECMWF), which is commonly used to provide boundary conditions for high resolution Limited Area Models (LAMs). We also highlight the usefulness of a global, relatively coarse, coupled atmosphere-ocean model for probabilistic forecasting and early warning of storm surge along complex coastlines.

Here we present a comparison of two standalone global NEMO \citep{gurvan_madec_2024} ocean model integrations covering the period 2015--2022, inclusively, with and without the atmospheric pressure effect, forced with the atmospheric reanalysis ERA5 \citep{hersbach20era5}. The model is identical to the ocean model component of the ECMWF IFS. See \citealt{mogensen12} and \citealt{mogensen17} for further details on the ocean model component of IFS. No astronomical tidal constituents are included in the model setup. This means that the water level given by the model is only the atmospheric contribution to the total water level. It is comparable to the residual water level due to atmospheric effects in the observations, i.e. the residual when subtracting the astronomical tides from the observed total water level, not taking non-linear interactions between tides and the atmospheric effects into account. 

The purpose of our study is two-fold. First, we want to investigate how well a coarse (roughly $0.25^\circ$) fully baroclinic ocean model can be expected to represent coastal and open-ocean water level variations associated with the passage of tropical and extratropical cyclones. Secondly, we want to assess how well a model without the inverse barometer effect (only Ekman effects) can be expected to perform. %This we do by diagnostically adding the inverse barometer effect to a control experiment with constant sea level pressure. 

Two experiments have been performed to assess two alternative ways of including the IB effect and its impact on the modeled sea surface elevation. This leaves us with three variables representing the water level,  referred to as \textit{noIB}, the control experiment with no IB effect, \textit{diaIB}, diagnostically adding the IB effect to the control experiment, and finally \textit{dynIB}, the water level from NEMO including the dynamically computed effect of the sea level pressure. Wind stress, and its effect on the water level, is included in all three variables.

Finally, the ensemble prediction system of the IFS (ENS) is used to perform a case study of the North Sea winter storm Xaver in December 2013. By examining the ensemble spread for the water level at two coastal locations in the North Sea, the impact of including the diagnostic IB effect is explored for probabilistic forecasting.

This article is structured as follows. In Section \ref{sec:obs_dataset} we present the dataset of observations used in the validation. In Section \ref{sec:model} the ocean model is described. In Section \ref{sec:results} we present statistics for the model results and the validation results. In Section \ref{sec:casestudies} a few case studies of storm surge events from within the hindcast period are evaluated in detail. Section \ref{sec:ec_ens_use} showcases how the results from this research can have direct implications in operational forecasting of storm surge. A summary and concluding remarks are found in Section \ref{sec:summary}.

%%%%%%%%%%%%%%%%%%%%%%%%%%%%%%%%%%%%%%%%%%%%%%%%%%%%%%%%%%%%%%%%%%%%%%%%%%%%%%%%%%%%%%%%%%%%%%%%%%%%%%%%%%%%%%%%%%%%%%%%%%%%
%%%%%%%%%%%%%%%%%%%%%%%%%%%%%%%%%%%%%%%%%%%%%%%%%%%%%%%%%%%%%%%%%%%%%%%%%%%%%%%%%%%%%%%%%%%%%%%%%%%%%%%%%%%%%%%%%%%%%%%%%%%%
\section{Observational dataset}\label{sec:obs_dataset}
\begin{figure}
    \centering
    \includegraphics[width=0.85\linewidth]{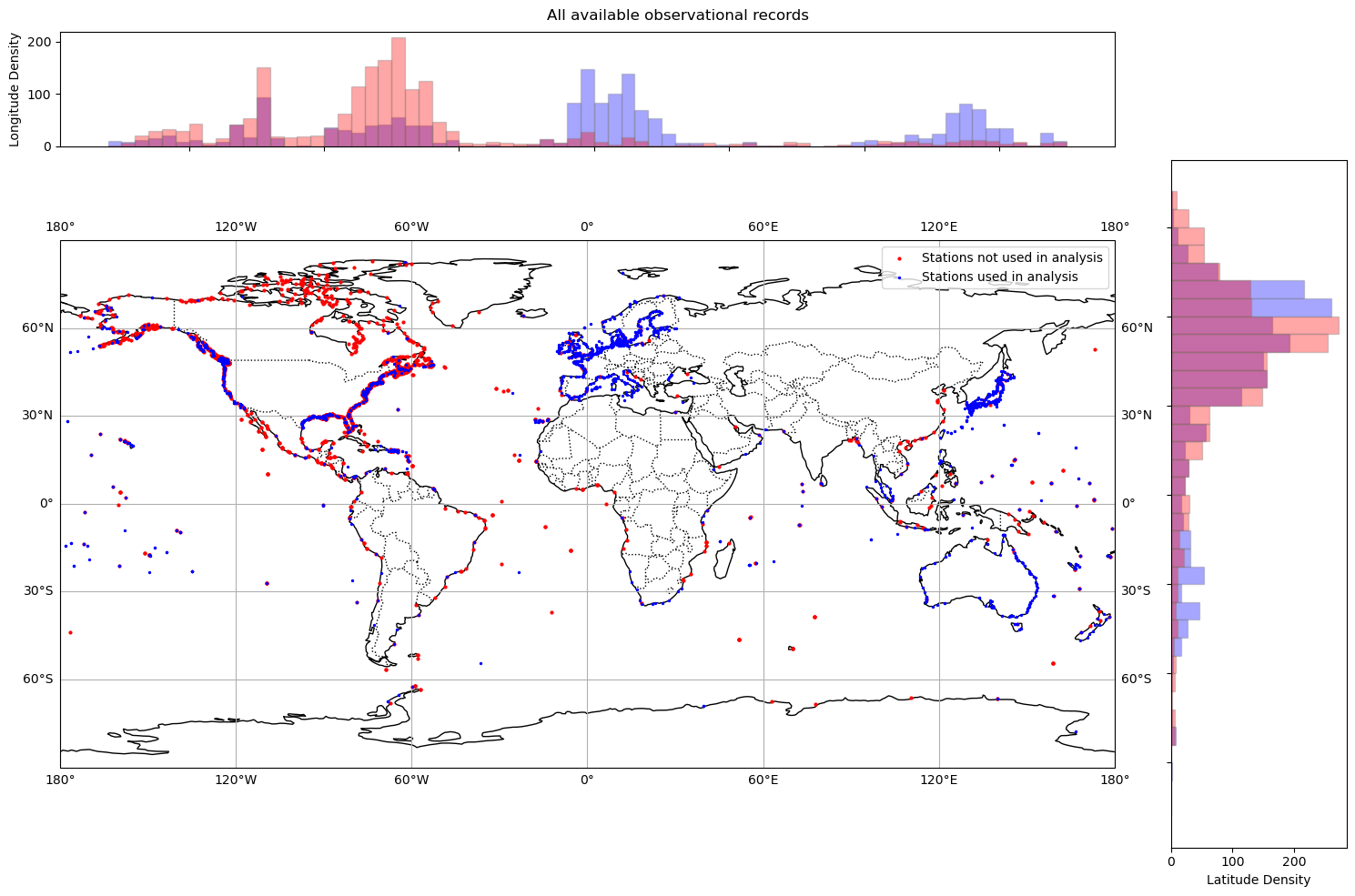}
    \caption{Positions of all available observation records in the GESLA3 dataset. The blue dots indicate the positions of those records that was used in this work, while the red dots were omitted either due to no overlapping time coverage between model and observation, or due to the record being located too far away from a model ocean point. The horizontal and vertical histograms indicate the geographical distribution in longitude and latitude directions of the records.}
    \label{fig:gesla3}
\end{figure}

The observation dataset used in this work is the Global Extreme Sea Level Analysis version 3 (GESLA3), see \citealt{haigh2023}. GESLA3 is a comprehensive and freely available dataset containing observation records from 36 different sources covering large parts of the globe. The dataset consists of 5119 stations with time series of observed total water level (TWL) on a common data format, covering the period from 1800 until 2020. It contains an aggregated total of 91,021 years for all the records. We note that some observation stations are contained in more than one of the sources used to compile the dataset, hence the total number of unique observation stations is less than the number of records. The locations and geographical distribution of the records are shown in Figure \ref{fig:gesla3}. 
To isolate the weather-induced effects on the water level, i.e. the \textit{residual water level}, from the TWL, we performed a tidal analysis of each record using 11 of the dominant tidal constituents\footnote{The tidal constituents used in the de-tiding are: M2, S2, N2, K1, M4, O1, MN4, Q1, P1, K2, MS4} in the pangeo-pytide python package, based on the FES2014 global ocean tide atlas \citep{lyard21}, and subtracted this tidal signal from the observed TWL. This is hereafter referred to as the observed residual water level or residual. Before the analysis, the temporal mean for both model time series and observation records is removed. We utilize all the records that overlap in time with our simulation period, located no more than two grid points away from an ocean grid point in the model. This reduces the number of records to be used in the analysis to around 1000 records. The blue dots in Figure \ref{fig:gesla3} indicate the locations and geographical distribution of the records used in our analysis, while the red dots indicate the records that was not used.
All observation records have been resampled to hourly time resolution through linear interpolation to facilitate the comparison with ocean model data. This was done after the tidal analysis was performed.

A feature of the GESLA3 dataset is that some stations appear as multiple records in the dataset, i.e., are provided from multiple sources, like, e.g., the national hydro-meteorological services and the Pan-European Copernicus Marine Service (CMEMS). These seem to occasionally be resampled or interpolated from their original temporal resolution to a different resolution, often hourly. When performing the tidal analysis on these records, and subtracting the tidal signal from the TWL, we are often left with a very noisy residual signal. This tends to clutter the model evaluation presented in Section \ref{sec:results}, and even if our brief investigation into the issue suggests that the number of ``noisy'' data points is small, this feature could have an impact on the general statistics when such records are included. More details and a visual example can be found in Appendix \ref{sec:app:error}.
Attempts to remove these outliers have not been successful, and we are therefore left with the choice of either omitting the records entirely, or leaving them in the dataset. We have chosen the latter.

%%%%%%%%%%%%%%%%%%%%%%%%%%%%%%%%%%%%%%%%%%%%%%%%%%%%%%%%%%%%%%%%%%%%%%%%%%%%%%%%%%%%%%%%%%%%%%%%%%%%%%%%%%%%%%%%%%%%%%%%%%%%
%%%%%%%%%%%%%%%%%%%%%%%%%%%%%%%%%%%%%%%%%%%%%%%%%%%%%%%%%%%%%%%%%%%%%%%%%%%%%%%%%%%%%%%%%%%%%%%%%%%%%%%%%%%%%%%%%%%%%%%%%%%%
\section{Model setup}\label{sec:model}
\subsection{Model description}
The NEMO ocean model \citep{gurvan_madec_2024} is here run with hourly atmospheric forcing from the ERA5 reanalysis \citep{hersbach20era5}. The domain is a global stretched tripolar ORCA025 grid with an average resolution of $0.25 \times 0.25^\circ$ which is refined threefold in the Equatorial band. The vertical is resolved with a $z$ coordinate with variable resolution. The upper level is $1\,\mathrm{m}$ thick. This setup is commonly referred to as \texttt{ORCA025\_Z75}.
The ocean model reads the 10 m wind vectors, temperature, humidity, heat and water fluxes from ERA5 every hour on the native TL639 model resolution, which translates into approximately $31~\mathrm{km}$ horizontal resolution. The ocean model also reads forcing from the ECWAM wave model (a component of ERA5) as described by \citealt{breivik2015}. 

\subsection{Numerical experiments}
We have designed three experiments to isolate the impact of the implementation of the IB effect within a wind-forced system (see Table \ref{tab:experiments}). All experiments cover the period 2015--2022 (inclusively, initialized from a warm start), and none of the runs includes astronomical tides.

Firstly, we perform an ocean-only integration with a constant MSLP (experiment id i5k6) as the control run. We then compute the inverse barometer effect from the ERA5 mean sea level pressure (MSLP) and output the total sea surface elevation (including the inverse barometer effect) as well as the prognostic elevation (only frictional and baroclinic effects). The sea surface elevation from these two experiments are hereafter referred to as \textit{noIB} for the run using static IB, and \textit{diaIB} for the run where we compute the inverse barometer effect diagnostically from the local MSLP. 
The control run is then compared against an otherwise identical ocean-only run where the ocean model is allowed to respond to a varying MSLP taken from ERA5 (experiment id i5k8). This run is referred to as the dynamical experiment (\textit{dynIB}). % whereas the former is referred to as the diagnostic experiment (\textit{diaIB}) . 
%The variables of interest in the control run are thus the sea surface elevation without the inverse barometer effect, referred to as \textit{noIB}, and the  sea surface elevation including the diagnostically computed inverse barometer effect, \textit{diaIB}. These are compared against the total sea surface elevation \textit{dynIB} in the alternative model setup with dynamical MSLP.

\begin{table}
    \centering
    \begin{tabular}{lm{4cm}lm{5cm}}
        \hline
        \textbf{Experiment} & \textbf{IB implementation} & \textbf{Key physical components} \\ \hline
        \textbf{noIB}  & None (i.e. constant MSLP) & Wind stress, Ekman transport, baroclinic response \\
        \textbf{diaIB}  & Diagnostic (post-process) & \textbf{noIB} components + diagnostic IB response \\
        \textbf{dynIB}  & Dynamical (at run-time) & Joint pressure-wind dynamics; dynamical IB response \\ \hline
    \end{tabular}
    \caption{Key components of the three experiments conducted to evaluate the two methods of including the IB effect.}
    \label{tab:experiments}
\end{table}

In Section \ref{sec:ec_ens_use}, we use the IFS ensemble prediction system \citep[ENS, e.g.][]{ens} to showcase how inclusion of the IB effect can impact the probabilistic forecast of storm surges. ENS is a coupled global forecasting system, which uses NEMO for its ocean model component, thus enabling us to explore how the impacts seen from the standalone NEMO studies translate into a coupled forecasting scenario.

Both the uncoupled and coupled (ENS) forecasts use NEMO version 4.0, which is a newer version than the 3.4 version of NEMO which is currently in operational use at ECMWF. The coupled model runs were based on a development version of IFS CY49R2 to incorporate the changes needed to run the coupled setup with NEMO V4.0. 

%%%%%%%%%%%%%%%%%%%%%%%%%%%%%%%%%%%%%%%%%%%%%%%%%%%%%%%%%%%%%%%%%%%%%%%%%%%%%%%%%%%%%%%%%%%%%%%%%%%%%%%%%%%%%%%%%%%%%%%%%%%%
%%%%%%%%%%%%%%%%%%%%%%%%%%%%%%%%%%%%%%%%%%%%%%%%%%%%%%%%%%%%%%%%%%%%%%%%%%%%%%%%%%%%%%%%%%%%%%%%%%%%%%%%%%%%%%%%%%%%%%%%%%%%
\section{Results}\label{sec:results}

%%%%%%%%%%%%%%%%%%%%%%%%%%%%%%%%%%%%%%%%%%%%%%%%%%%%%%%%%%%%%%%%%%%%%%%%%%%%%%%%%%%%%%%%%%%%%%%%%%%%%%%%%%%%%%%%%%%%%%%%%%%%
\subsection{Model statistics}\label{sec:res:modstats}
\begin{figure}
  \centering
  \subfloat[Diagnostic]{\includegraphics[width=0.5\textwidth]{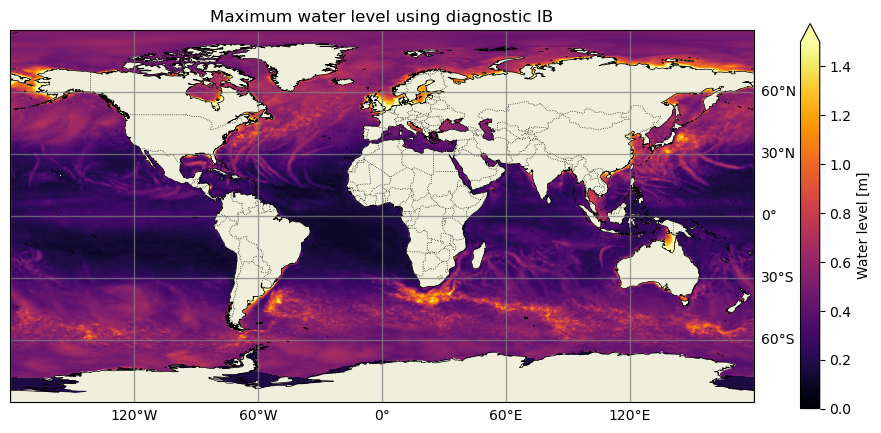}\label{fig:map_max_dia}}
  \subfloat[Difference]{\includegraphics[width=0.5\textwidth]{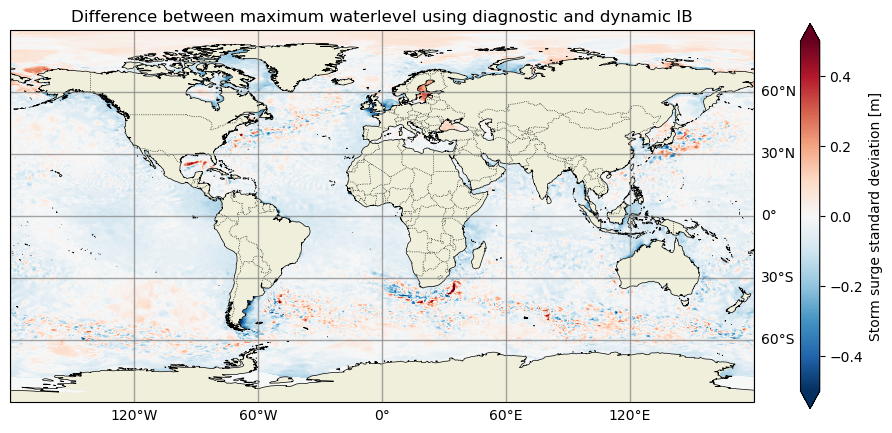}\label{fig:map_max_diff}}
  \caption{The maximum water level in the stand-alone storm surge hindcast with the IB effect included diagnostically as a post-process is displayed in panel (a). Panel (b) shows the difference between the diagnostic and dynamical inclusion of the IB effect. The red areas indicate where the diagnostic inclusion has the largest maximum water level, and the blue areas where the dynamical inclusion give the largest maximum water level.}
  \label{fig:map_max}
\end{figure}

The map in Figure \ref{fig:map_max_dia} display the maximum values for water level for each grid point for the global domain of the NEMO model over the simulation period. The figure show the maxima for the \textit{diaIB} run, and the difference between the \textit{diaIB} and \textit{dynIB} runs.
The map of maxima in Figure \ref{fig:map_max} is a good indication of the areas around the globe that are prone to high water level due to storm surge. The Figure also highlight some of the well known dynamically active regions with large ocean currents, like e.g. the Gulf Stream east of the US East Coast, the Agulhas Current south and east of Africa and the Kuroshio Current east of Japan.
Due to the fact that high water level usually only poses a substantial risk to life and property when it occurs along a coast, we note a few coastal areas with large maxima for further investigation. These areas include the European North West Shelf (NWS), the southeast coast of South America, the southern and eastern coasts of the US and the coast of China.
In addition, when examining the maxima in Figure \ref{fig:map_max}, we can clearly see the imprints of the tracks of many of the major tropical storms, hurricanes and typhoons during the simulation period.

%%%%%%%%%%%%%%%%%%%%%%%%%%%%%%%%%%%%%%%%%%%%%%%%%%%%%%%%%%%%%%%%%%%%%%%%%%%%%%%%%%%%%%%%%%%%%%%%%%%%%%%%%%%%%%%%%%%%%%%%%%%%
\subsection{Model evaluation and validation}\label{sec:res:valid}
The control experiment with diagnostic inclusion of the IB effect (\textit{diaIB}) and the alternative (\textit{dynIB}) is compared with the GESLA3 observations. As a baseline, we also compare against the control experiment without inclusion of the IB effect (\textit{noIB}).
We focus the evaluation on the Mean Absolute Error (MAE) and the Root Mean Squared Error (RMSE), together with scaled versions of the two ($\mathrm{MAE}/\sigma$ and $\mathrm{RMSE}/\sigma$, where $\sigma$ is the standard deviation of the observed residual water level for each station). The values of MAE and RMSE for each record is scaled by the standard deviation of the observed residual water level for that record to provide a measure of the relative error, since the magnitude of the absolute error tends to be correlated with the magnitude of the residual, to allow us to compare the model performance across different records with different typical amplitudes. This is in line with the approach by \citealt{kristensen2024}, and the formulas can be found in Appendix B.

The MAE and RMSE [m] for the water level of \textit{noIB}, \textit{diaIB} and \textit{dynIB} are listed in Table \ref{tab:stats_abs}. We focus on the median of the error metrics across the records, but also include the mean values. The median is chosen as it is a robust estimate of central tendency and thus suitable for analysis given the large outliers found in some of the observation records, as explained in Section \ref{sec:obs_dataset}.
We include the MAE and RMSE for the entire hindcast period, denoted by the subscript \textit{all}, and also values where we mask the data when the observed absolute values of the residual is less than one and two standard deviations from the mean, denoted by the subscripts \textit{$\pm 1\mathrm{std}$} and {$\pm 2\mathrm{std}$}. This is done to emphasize the cases and time periods when the contribution from the atmospheric effects to the total water level is significant to focus on the model's ability to capture anomalously low and high water level events (the latter being what we usually think of as storm surges).
Similarly, the relative (scaled) values $\mathrm{MAE}/\sigma$ and $\mathrm{RMSE}/\sigma$, are presented in Table \ref{tab:stats_rel}. 

The absolute values of both MAE and RMSE in Table \ref{tab:stats_abs} clearly show that the errors increase when masking out the low absolute values. Also, it is evident that \textit{diaIB} has slightly lower errors, $1\,\mathrm{cm}$ less for both MAE and RMSE, compared to the \textit{dynIB}. When including all data, the best performing experiment, \textit{diaIB}, has a MAE of $10\,\mathrm{cm}$ and RMSE of $13\,\mathrm{cm}$ averaged over the entire global model domain and simulation period. When masking the data for observations within $\pm \sigma$, the MAE and RMSE for the \textit{diaIB} increase to $12$ and $15\,\mathrm{cm}$ respectively. And for the masking criteria of observations within $\pm 2\sigma$, the MAE and RMSE is $15$ and $18\,\mathrm{cm}$ for \textit{diaIB}, respectively. This is approximately $30\%$ less than the MAE and RMSE of \textit{noIB}. This clearly shows that, unsurprisingly, to properly simulate the water level during an extreme storm surge event, the IB effect must be included.

\begin{table}
    \centering
    \begin{tabular}{lccc}
        \hline
         & \textbf{No IB} & \textbf{Diagnostic IB} & \textbf{Dynamic IB} \\ \hline
        $\textbf{MAE}_{all}$ & 11 (15) & 10 (15) & 11 (17) \\ %\hline
        $\textbf{RMSE}_{all}$ & 14 (20) & 13 (18) & 14 (21) \\ \hline
        $\textbf{MAE}_{\pm1std}$ & 16 (25) & 12 (21) & 13 (23) \\ %\hline
        $\textbf{RMSE}_{\pm1std}$ & 18 (28) & 15 (25) & 16 (27) \\ \hline
        $\textbf{MAE}_{\pm2std}$ & 22 (33) & 15 (27) & 16 (28) \\ %\hline
        $\textbf{RMSE}_{\pm2std}$ & 25 (37) & 18 (32) & 19 (33) \\ \hline
    \end{tabular}
    \caption{Median values (arithmetic means in parentheses) for Mean Absolute Error (MAE) and Root Mean Square Error (RMSE) for all records when comparing the NEMO hindcast and GESLA3 observations. MAE and RMSE are shown for all data, for when masking the values less than $\pm 1$ and $\pm 2$ standard deviations. Values are given in centimeters.}
    \label{tab:stats_abs}
\end{table}

\begin{table}
    \centering
    \begin{tabular}{lccc}
        \hline
         & \textbf{No IB} & \textbf{Diagnostic IB} & \textbf{Dynamic IB} \\ \hline
        $\textbf{MAE}_{all}/\sigma$ & 0.68 (0.76) & 0.60 (0.71) & 0.71 (0.83) \\ %\hline
        $\textbf{RMSE}_{all}/\sigma$ & 0.87 (0.96) & 0.77 (0.89) & 0.89 (1.03) \\ \hline
        $\textbf{MAE}_{\pm1std}/\sigma$ & 1.03 (1.12) & 0.77 (0.92) & 0.84 (1.0) \\ %\hline
        $\textbf{RMSE}_{\pm1std}/\sigma$ & 1.20 (1.32) & 0.94 (1.12) & 1.04 (1.22) \\ \hline
        $\textbf{MAE}_{\pm2std}/\sigma$ & 1.53 (1.63) & 1.0 (1.24) & 1.06 (1.28) \\ %\hline
        $\textbf{RMSE}_{\pm2std}/\sigma$ & 1.67 (1.83) & 1.20 (1.46) & 1.27 (1.52) \\ \hline
    \end{tabular}
    \caption{Same as Table \ref{tab:stats_abs}, but with values are given as fractions of standard deviations (absolute values scaled by the standard deviation for the observation for each record). Arithmetic means are given in parentheses.}
    \label{tab:stats_rel}
\end{table}

\begin{figure}[ht]
  \centering
  \subfloat[All data]{\includegraphics[width=0.32\textwidth]{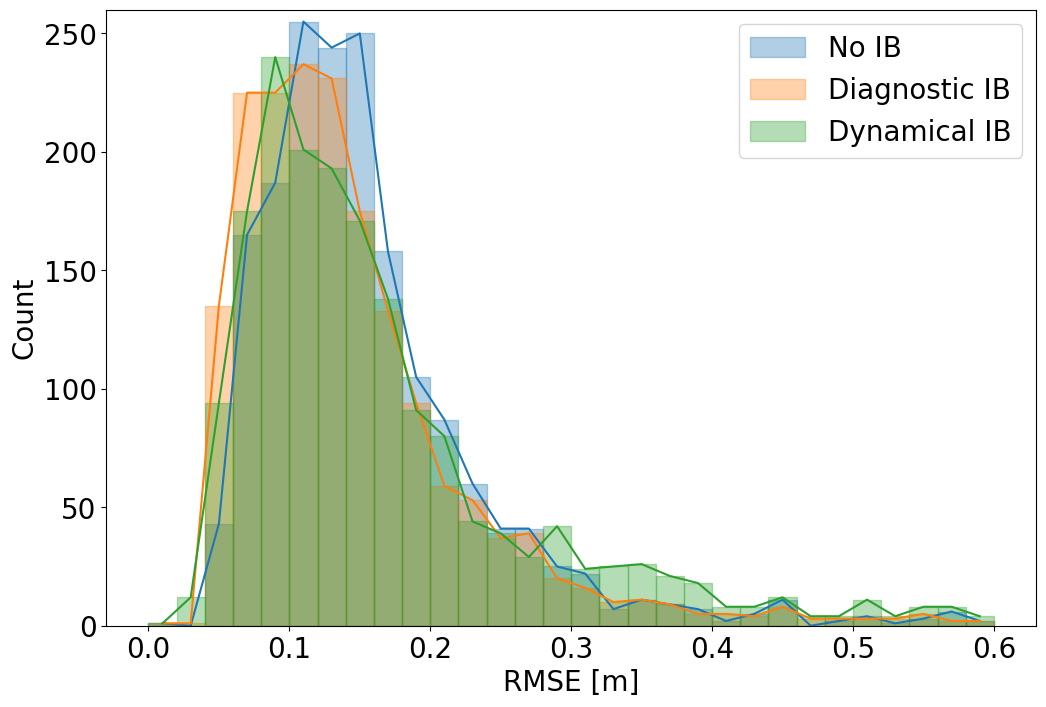}\label{fig:errdist_all_abs}}
  % \hfill
  \subfloat[$|\mathrm{obs}| >  \sigma$]{\includegraphics[width=0.32\textwidth]{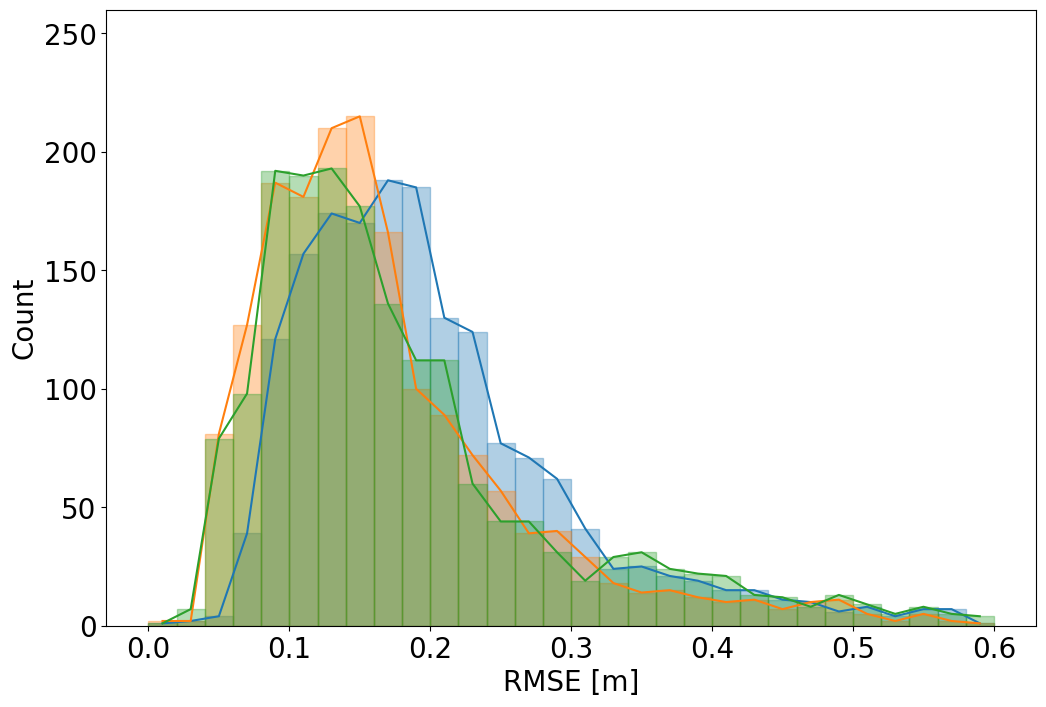}\label{fig:errdist_pm1_abs}}  
  % \hfill
  \subfloat[$|\mathrm{obs}| > 2\sigma$]{\includegraphics[width=0.32\textwidth]{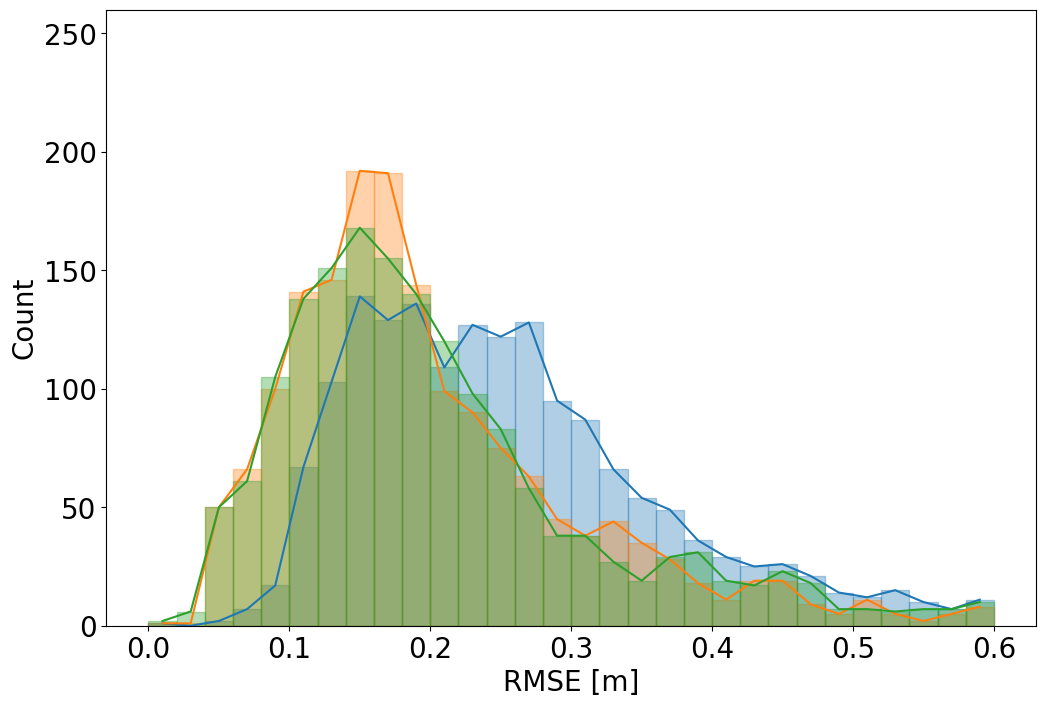}\label{fig:errdist_pm2_abs}}  
  \hfill
  \subfloat[All data]{\includegraphics[width=0.32\textwidth]{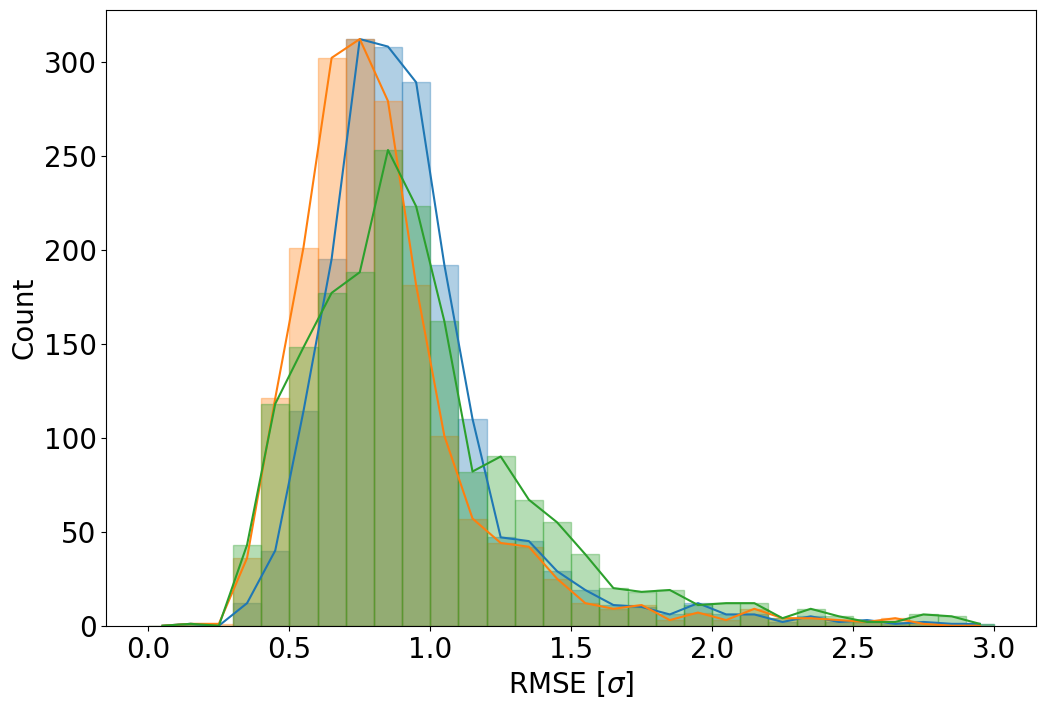}\label{fig:errdist_all_scaled}}
  % \hfill
  \subfloat[$|\mathrm{obs}| > \sigma$]{\includegraphics[width=0.32\textwidth]{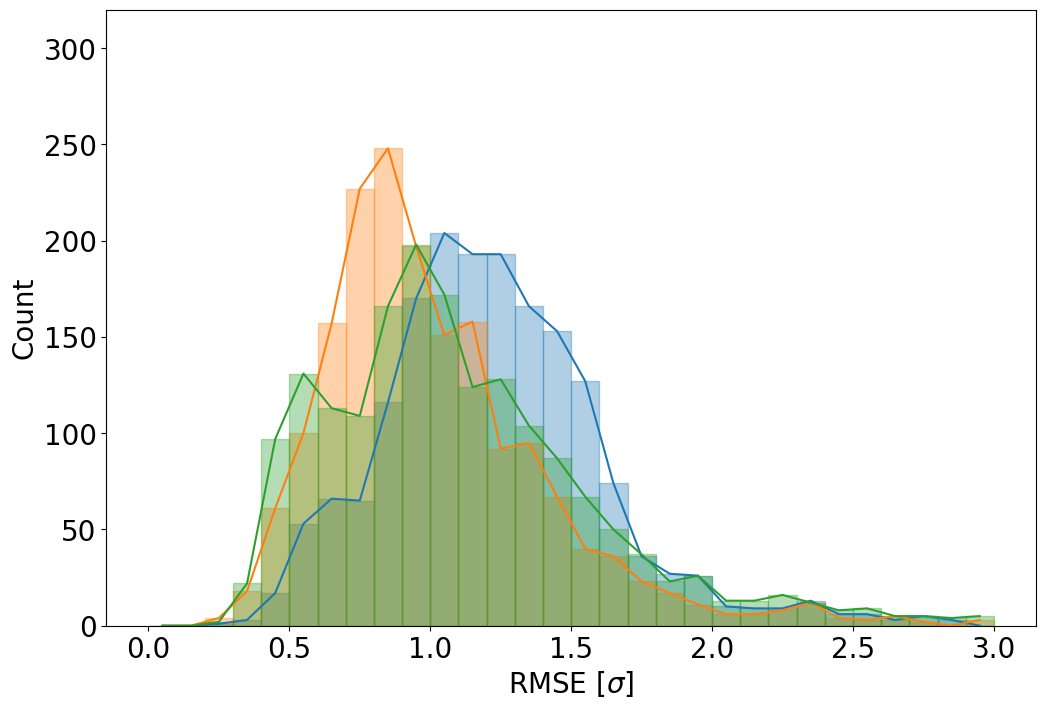}\label{fig:errdist_pm1_scaled}}  
  % \hfill
  \subfloat[$|\mathrm{obs}| > 2\sigma$]{\includegraphics[width=0.32\textwidth]{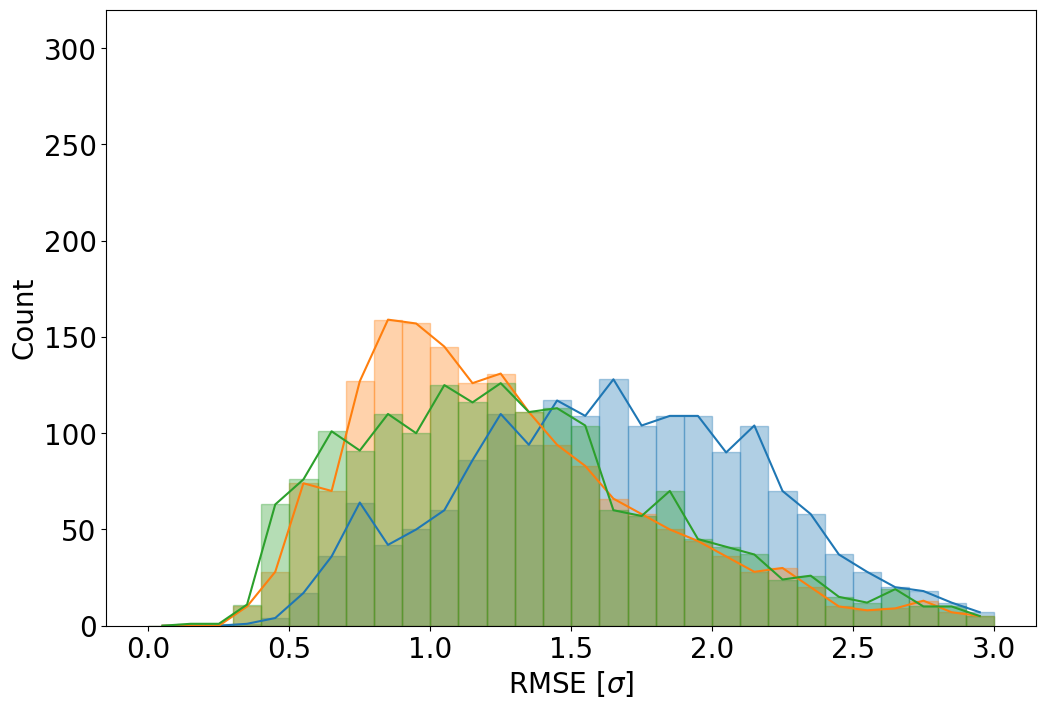}\label{fig:errdist_pm2_scaled}}  
  \caption{Distribution of the Root Mean Square Error (RMSE) for all records for the three different representations of the atmospheric effects on the water level. The top panels (a-c) present the RMSE in meters, with a bin size of 2 cm. The bottom panels (d-f) show the RMSE scaled by the standard deviation ($\sigma$) of each observation record, using a bin size of $0.1\sigma$. Panels (a) and (d) present the RMSE distribution for the complete dataset. Panels (b) and (e) illustrate the RMSE for observed residual water level absolute values more than one standard deviation ($|\mathrm{obs}| > \sigma$). Panels (c) and (f) show the RMSE for observed residual water level absolute values exceeding two standard deviations ($|\mathrm{obs}| > 2\sigma$). The $y$-axis indicates the number of occurrences (records).}
  \label{fig:errdist}
\end{figure}

However, Figure \ref{fig:errdist} shows that the statistical distribution of the RMSE has a large variation, with the majority of the errors in the range $4-30\,\mathrm{cm}$ and $[0.3-1.6]\sigma$, when considering all data. It is evident that all the distributions, except the relative \textit{noIB}, are skewed to the right. This is as expected based on the mean and median values in Table \ref{tab:stats_abs} and \ref{tab:stats_rel} where the mean value is generally larger than the median. The width of the distributions increases with stricter masking criteria.

\begin{figure}[ht]
  \centering
  \includegraphics[width=1\textwidth]{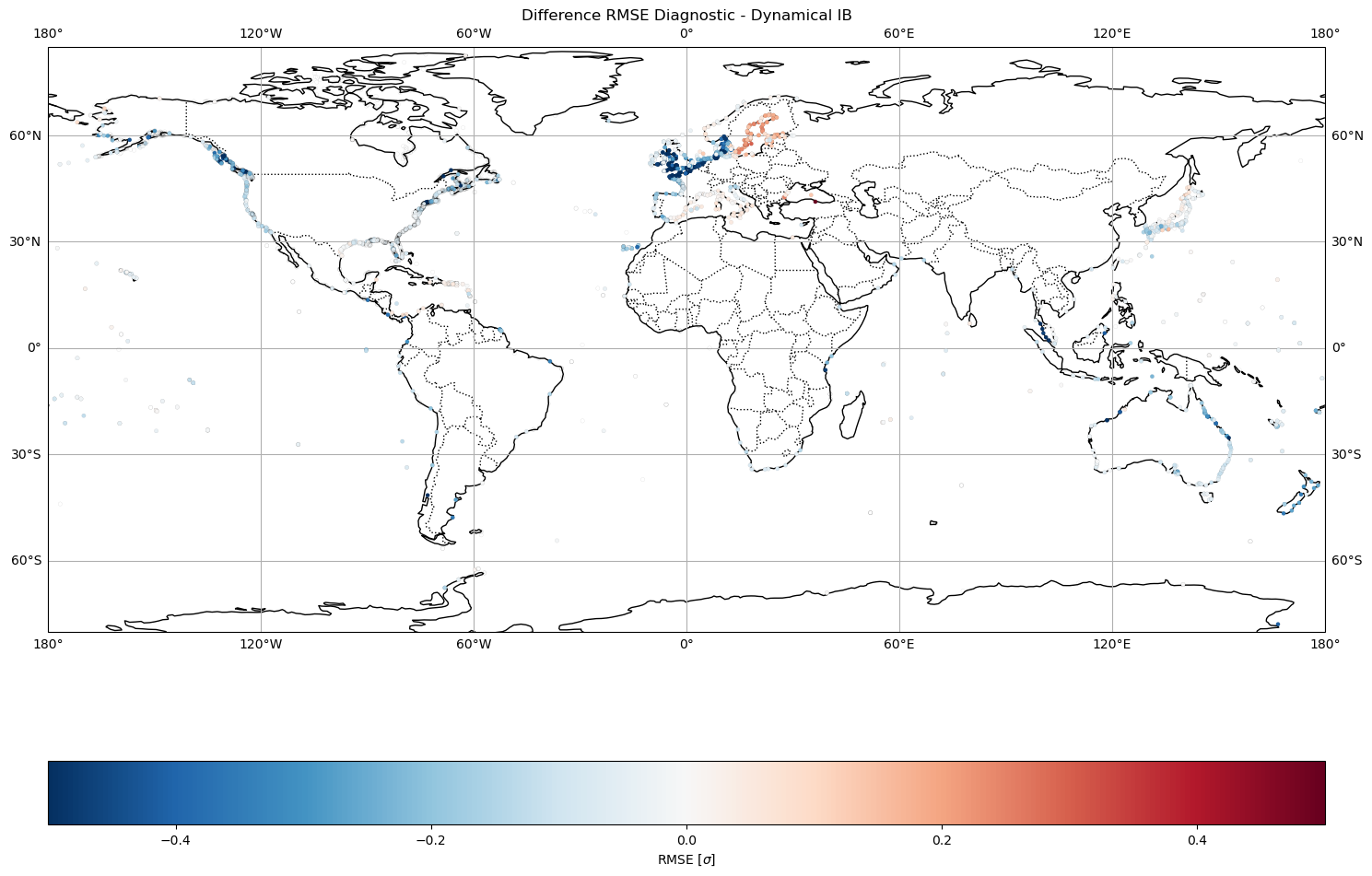}
  \caption{Map of difference in RMSE for all records showing geographical differences. The circles indicate the position of the record, and the color indicate the difference in relative RMSE in number of standard deviations between the diagnostic and dynamical IB. Blue colors indicate that the diagnostic IB performs better than the dynamical IB, and red vice versa.}
  \label{fig:map_rmse_diff_rel}
\end{figure}

\begin{figure}
    \centering
    \includegraphics[width=0.75\linewidth]{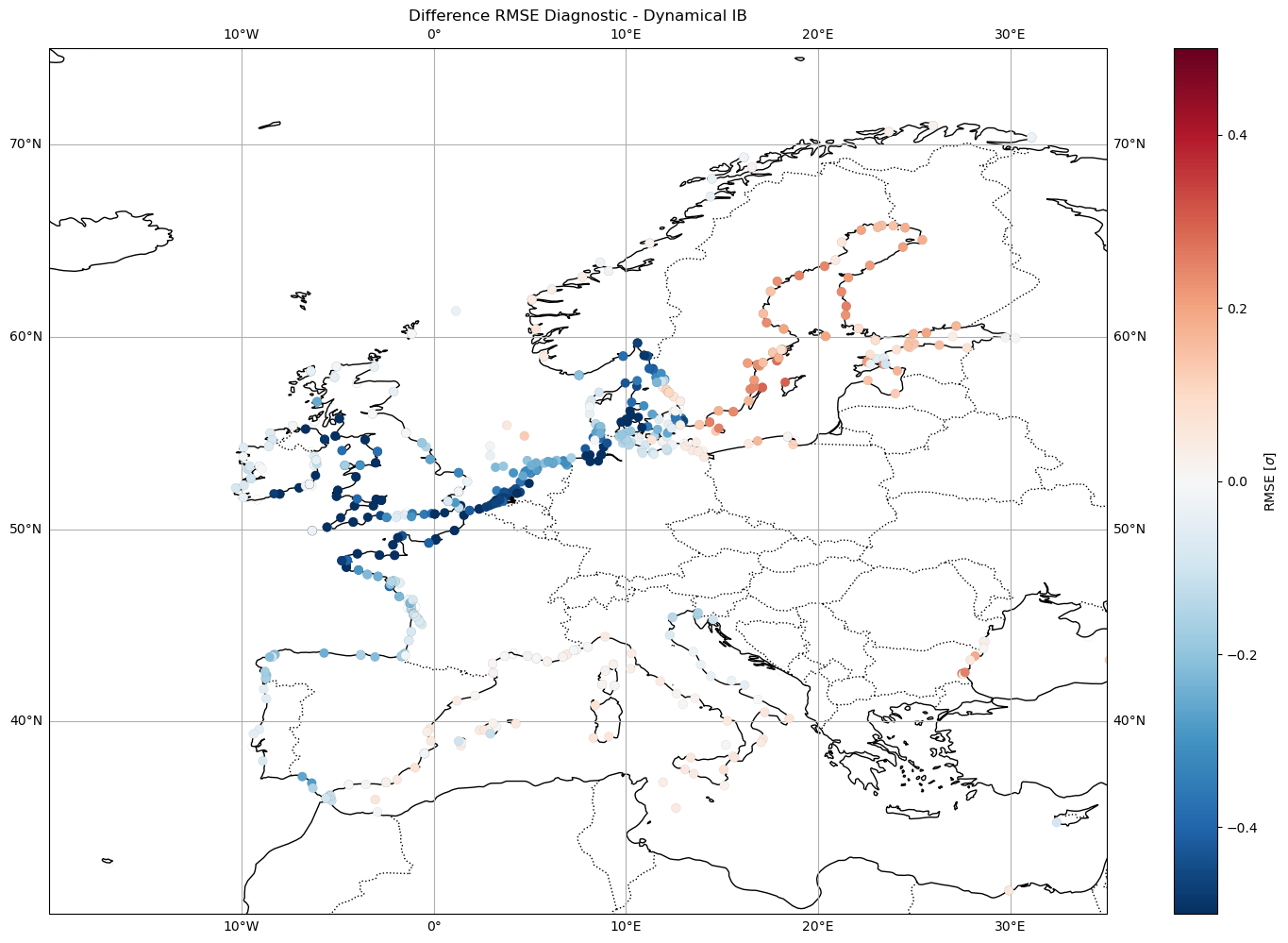}
    \caption{Same as Figure \ref{fig:map_rmse_diff_rel}, but zoomed to the European region to highlight the difference between the European North West Shelf and Baltic Sea regions. Blue colors indicate that the diagnostic IB performs better than the dynamical IB, and red vice versa.}
    \label{fig:map_rmse_diff_rel_eur}
\end{figure}

Additionally, the map in Figure \ref{fig:map_rmse_diff_rel} clearly shows the regional differences in performance between the two experiments. The difference in relative error, displayed in Figure \ref{fig:map_rmse_diff_rel}, highlights the European NWS and the Baltic Sea as two regions (see also Figure \ref{fig:map_rmse_diff_rel_eur}) where the differences between the two experiments are large with  \textit{diaIB} outperforming \textit{dynIB} on the European NWS, and vice versa in the Baltic Sea. Most other areas exhibit relatively small differences, with a few scattered exceptions mostly in favor of the \textit{diaIB}.
It is unclear why the \textit{dynIB} experiment seems to outperform the diagnostic experiment in the semi-enclosed water body of the Baltic Sea. However, as we see a similar, albeit weaker, tendency in the Mediterranean Sea and the Black Sea, we speculate that the narrow connections to the open ocean limits the efficacy of the inverse barometer effect. These water bodies also have very small tidal amplitudes for the same reason, see e.g., \citealt{medvedev2016}. In other words, although the atmospheric pressure variations set up a pressure gradient, the narrows connecting the Baltic Sea \citep{weisse2021}, the Mediterranean Sea and the Black Sea (respectively the Danish Straits, the Strait of Gibraltar and the Bosphorus) are not wide enough for water to adjust on the synoptic time scales of hours to at most a few days. \cite{gomis2008} show that for the Mediterranean, the Strait of Gibraltar acts as a choking point and bottleneck where the Atlantic and Mediterranean have independent modes of variability due to the restriction of flow.
When the atmospheric pressure averaged over the entire confined sea area changes, this will lead to a change in the average water level in the \textit{diaIB}, whereas in reality and in the \textit{dynIB}, these changes in water level are mostly local, leaving the water level over the confined sea area mostly unchanged during the passage of extratropical systems. This effect becomes more pronounced when the dimensions of the confined water body coincides with the spatial extent of the weather systems.

While our global NEMO configuration employs a structured ORCA025 grid ($\sim31~\mathrm{km}$ resolution), it provides a relatively uniform coastal resolution similar to the 20 km coastal refinement used in the unstructured MOG2D-G mesh of \cite{carrere2003}. Consistent with their findings at high latitudes, our results show that including the IB effect (whether diagnostic or dynamic) is crucial for reducing errors in energetic regions. However, a key distinction in our study is that wind stress is included in all experiments (\textit{noIB}, \textit{diaIB} and \textit{dynIB}), allowing us to isolate the specific impact of the pressure implementation method.%, a nuance further explored in semi-enclosed basins where dynamic responses are restricted.

%%%%%%%%%%%%%%%%%%%%%%%%%%%%%%%%%%%%%%%%%%%%%%%%%%%%%%%%%%%%%%%%%%%%%%%%%%%%%%%%%%%%%%%%%%%%%%%%%%%%%%%%%%%%%%%%%%%%%%%%%%%%
%%%%%%%%%%%%%%%%%%%%%%%%%%%%%%%%%%%%%%%%%%%%%%%%%%%%%%%%%%%%%%%%%%%%%%%%%%%%%%%%%%%%%%%%%%%%%%%%%%%%%%%%%%%%%%%%%%%%%%%%%%%%
\section{Case studies}\label{sec:casestudies}
In addition to the general statistics, we now look at a few extratropical storm surge events on the European NWS, as well as tropical cyclones. The selected cases are from the areas of maximum storm surge in Figure \ref{fig:map_max}.

%%%%%%%%%%%%%%%%%%%%%%%%%%%%%%%%%%%%%%%%%%%%%%%%%%%%%%%%%%%%%%%%%%%%%%%%%%%%%%%%%%%%%%%%%%%%%%%%%%%%%%%%%%%%%%%%%%%%%%%%%%%%
\subsection{European/North Sea surges of January and February 2020}\label{sec:case:northsea}
January and February 2020 was a particularly active storm period in Northern Europe with several large storm surges. A series of powerful low pressure weather systems passed through Northern Europe from west to east, creating prolonged periods of high water level along the coast, in particular around the North Sea basin. In Norway, two storms during this time period were classified as ``extreme weather events'' and consequently were given names. These were the storms Didrik on 15 January 2020 \citep{skjerdal:etal:2020}, and Elsa 10--11 February 2020 \citep{selberg:etal:2020}. The storm Elsa was named Ciara in the United Kingdom (UK), see \citealt{jardine2023}, but we choose here to refer to it as Elsa.

\subsubsection{Storm Didrik}
The storm Didrik hit Southern Norway on 15 January 2020 \citep{skjerdal:etal:2020} with a forecast warning of high water level along the coast. The low pressure system, with a central pressure as low as $945 \,\mathrm{hPa}$, formed off the southern tip of Greenland on 12 January and moved eastward between Iceland and the UK the following days. It brought with it strong westerly to south-westerly winds in the northern North Sea, which in addition to the inverse barometer effect itself resulted in a large transport of water towards the western Norwegian coast. The maximum total water level observed was among the top 10 highest for the stations from Stavanger in the south west to Ålesund on the north-western coast \citep{kristensen2024}. 

\begin{figure}
    \centering
    \includegraphics[width=0.75\linewidth]{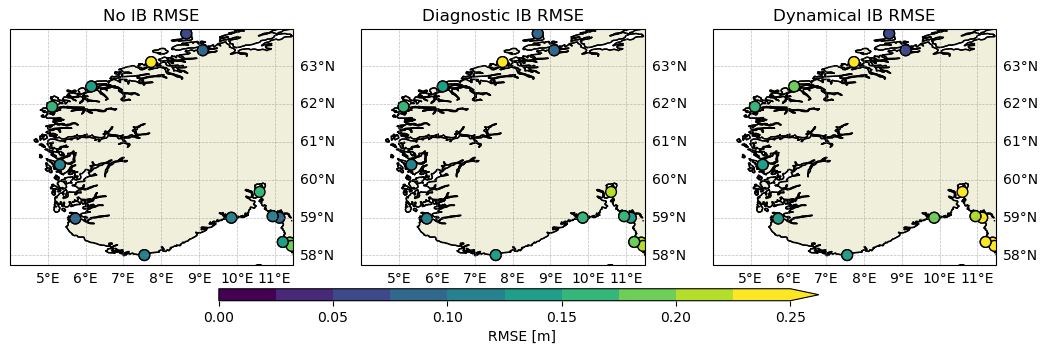}
    \caption{Maps of southern Norway showing the average RMSE for the daily maximum storm surge during the storm Didrik for the time period 14th-15th of January 2020.}
    \label{fig:didrik_map_err}
\end{figure}

\begin{figure}
    \centering
    \includegraphics[width=0.95\linewidth]{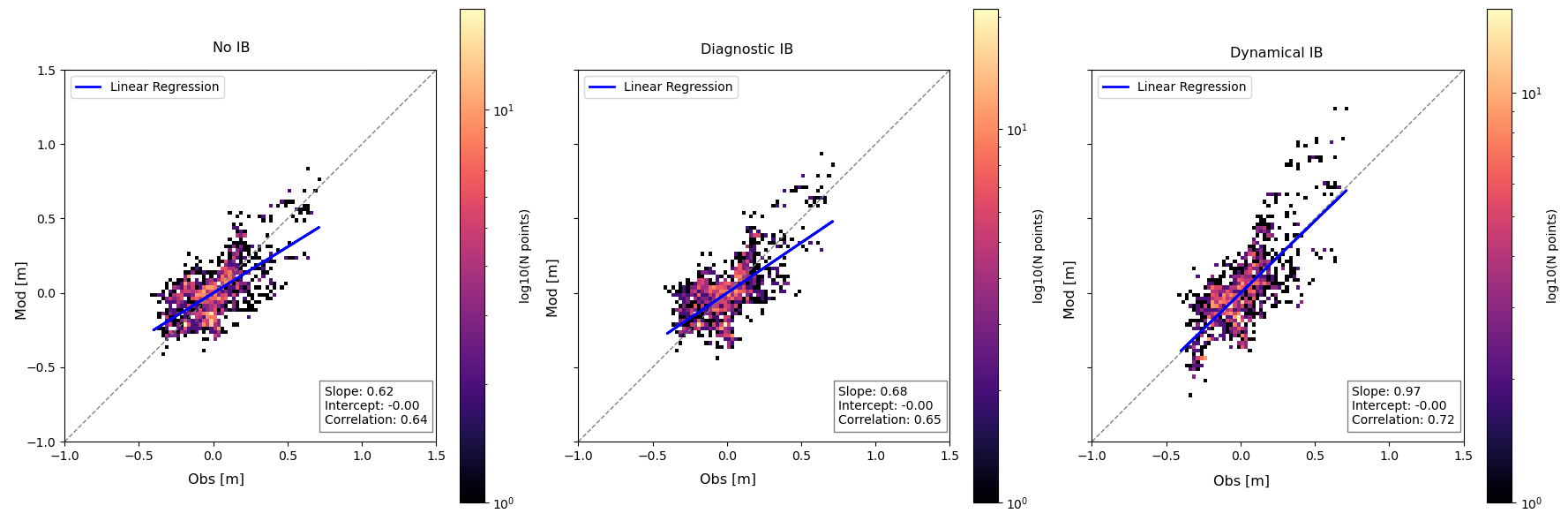}
    \caption{Heatmap scatterplot of the observed and modeled residual water levels for the positions shown in the maps in Figure \ref{fig:didrik_map_err}.}
    \label{fig:didrik_heatmap_scatter}
\end{figure}

\begin{figure}[ht]
  \centering
  \subfloat[Oscarsborg (NO)]{\includegraphics[width=0.5\textwidth]{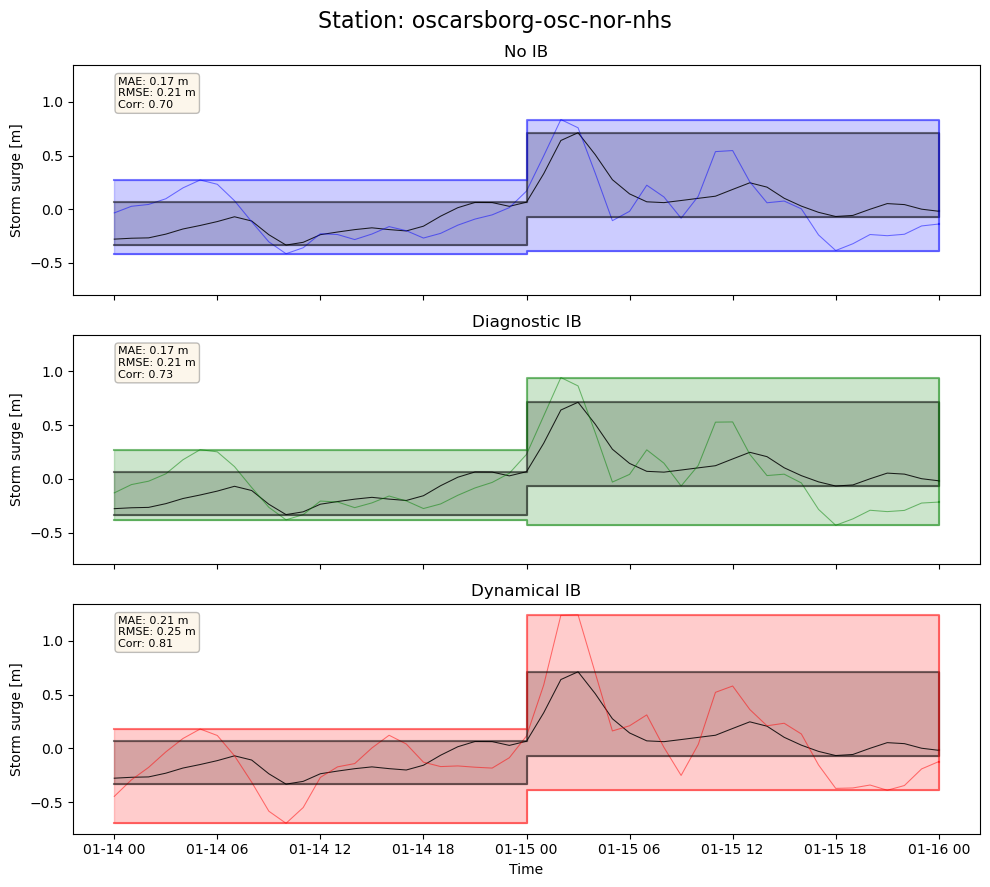}\label{fig:}}
  \subfloat[Bergen (NO)]{\includegraphics[width=0.5\textwidth]{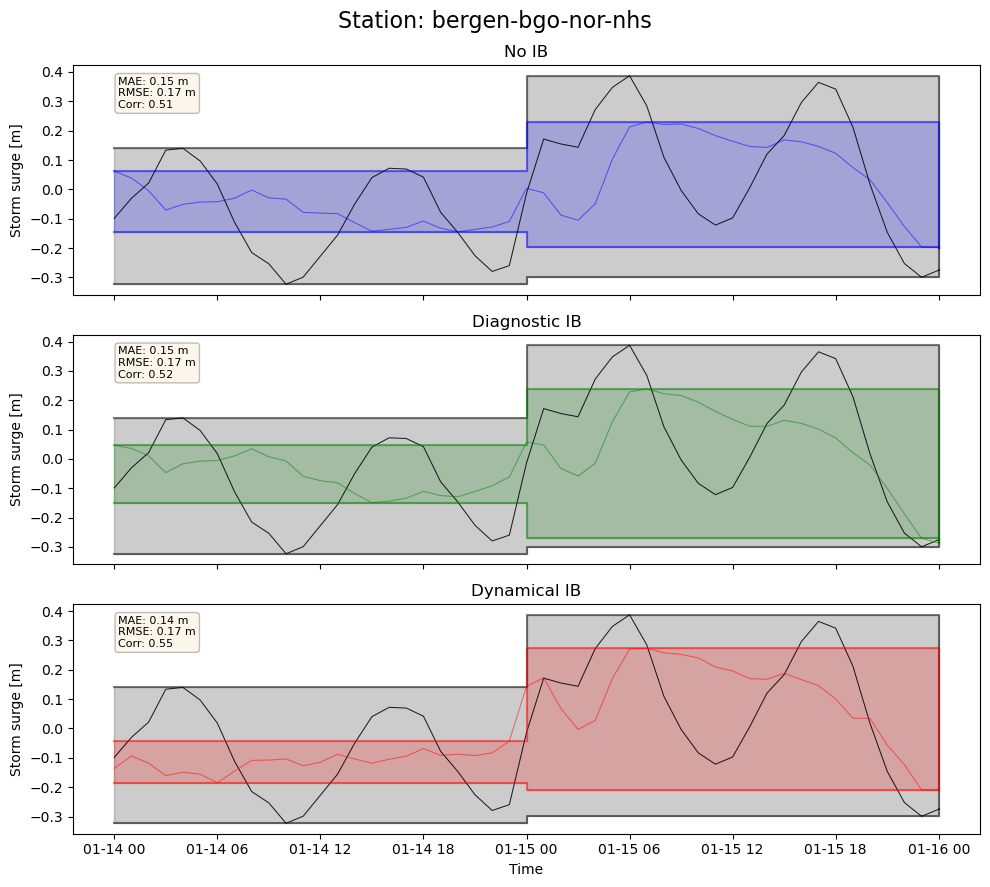}\label{fig:}}
  \caption{Time series for the Norwegian stations Oscarsborg and Bergen during the storm Didrik. The shaded areas indicate the range between daily minimum and maximum values, while the lines indicate the hourly values for model and observation. The textbox contains the MAE, RMSE and correlation values for the depicted time period for each station an parameter. The vertical reference for the time series is the mean sea level for the shown period.}
  \label{fig:didrik_timeseries}
\end{figure}

The maps in Figure \ref{fig:didrik_map_err} show the positions of the observations in southern Norway, colored by the average RMSE of the daily maximum values for storm surge. The differences between the experiments are very small, and only clearly visible for the locations furthest to the east, where the \textit{dynIB} has larger errors than \textit{noIB} and \textit{diaIB} due to too large amplitudes. This is also in line with the pattern in Figure \ref{fig:map_rmse_diff_rel_eur}. Time series for four selected stations in southern Norway are shown in Figure \ref{fig:didrik_timeseries}, where the Oscarsborg station (panel (a)) located in the Oslofjord in southeastern Norway, most clearly demonstrates how the amplitude at the maximum storm surge is overestimated.
The heatmap scatterplots in Figure \ref{fig:didrik_heatmap_scatter} suggest that the \textit{dynIB} is the best-performing experiment on average during storm Didrik with a correlation of $0.72$ compared to $0.65$ and $0.64$ for the \textit{diaIB} and \textit{noIB} respectively. However, it is quite clear that the maximum values of the modeled storm surge is closer to the observed surge for both \textit{noIB} and \textit{diaIB} compared to \textit{dynIB}, indicating that the overestimation of the storm surge at the south- and easternmost positions (as seen in panel (a) %and (b) 
in Figure \ref{fig:didrik_timeseries}) in the \textit{dynIB} experiment compensates for the underestimation that can be seen in all experiments in the western and northern stations.

\subsubsection{Storm Elsa}
\begin{figure}
    \centering
    \includegraphics[width=0.75\linewidth]{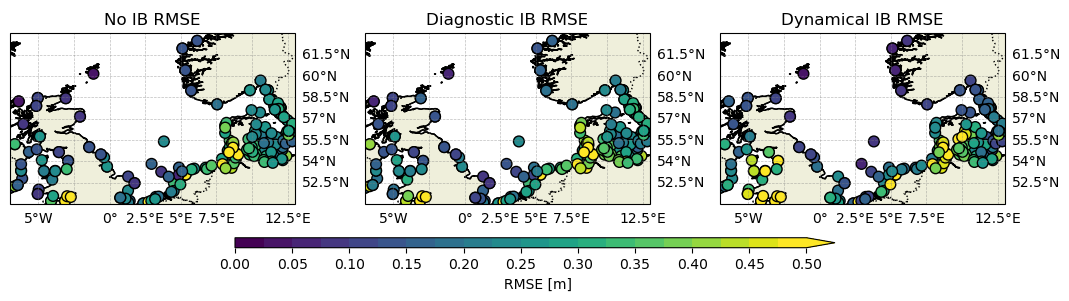}
    \caption{Maps of the North Sea area showing the average RMSE for the daily maximum storm surge during the storm Elsa for the time period 9th-12th of February 2020.}
    \label{fig:elsa_ciara_map_err}
\end{figure}

\begin{figure}
    \centering
    \includegraphics[width=0.95\linewidth]{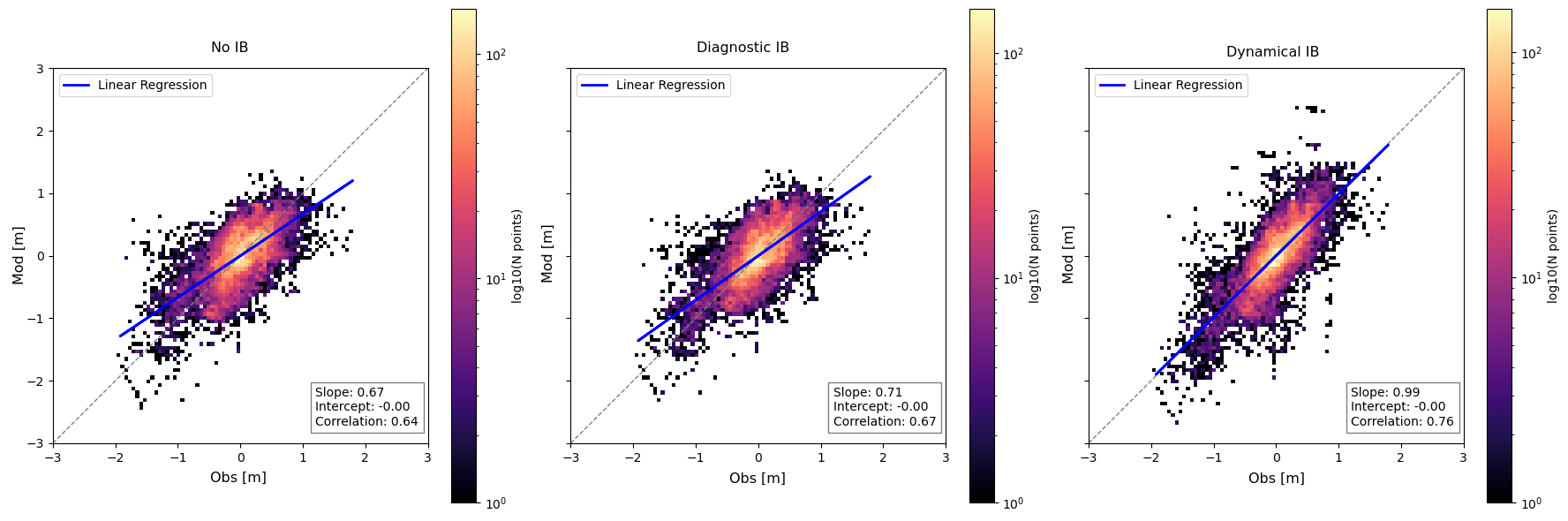}
    \caption{Heatmap scatterplot of the observed and modeled residual water levels for the positions shown in the maps in Figure \ref{fig:elsa_ciara_map_err}.}
    \label{fig:elsa_ciara_heatmap_scatter}
\end{figure}

\begin{figure}[ht]
  \centering
  \subfloat[Cuxhaven (GER)]{\includegraphics[width=0.5\textwidth]{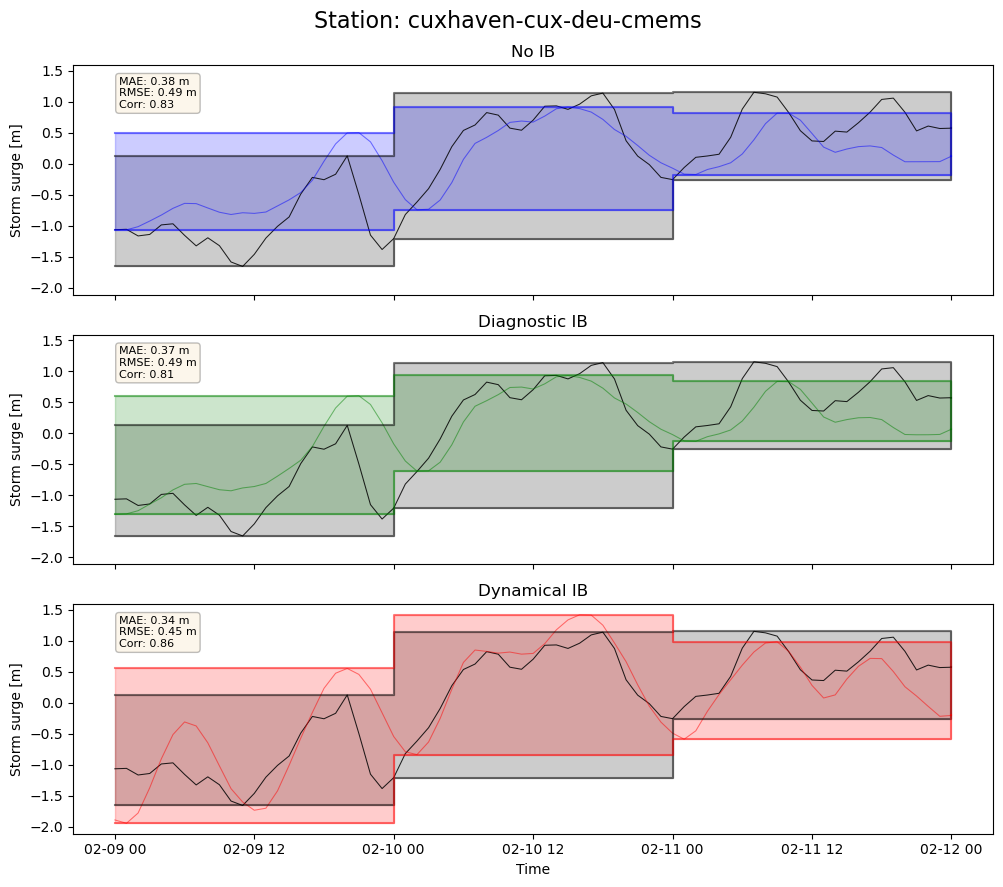}\label{fig:}}  
  \subfloat[Bergen (NO)]{\includegraphics[width=0.5\textwidth]{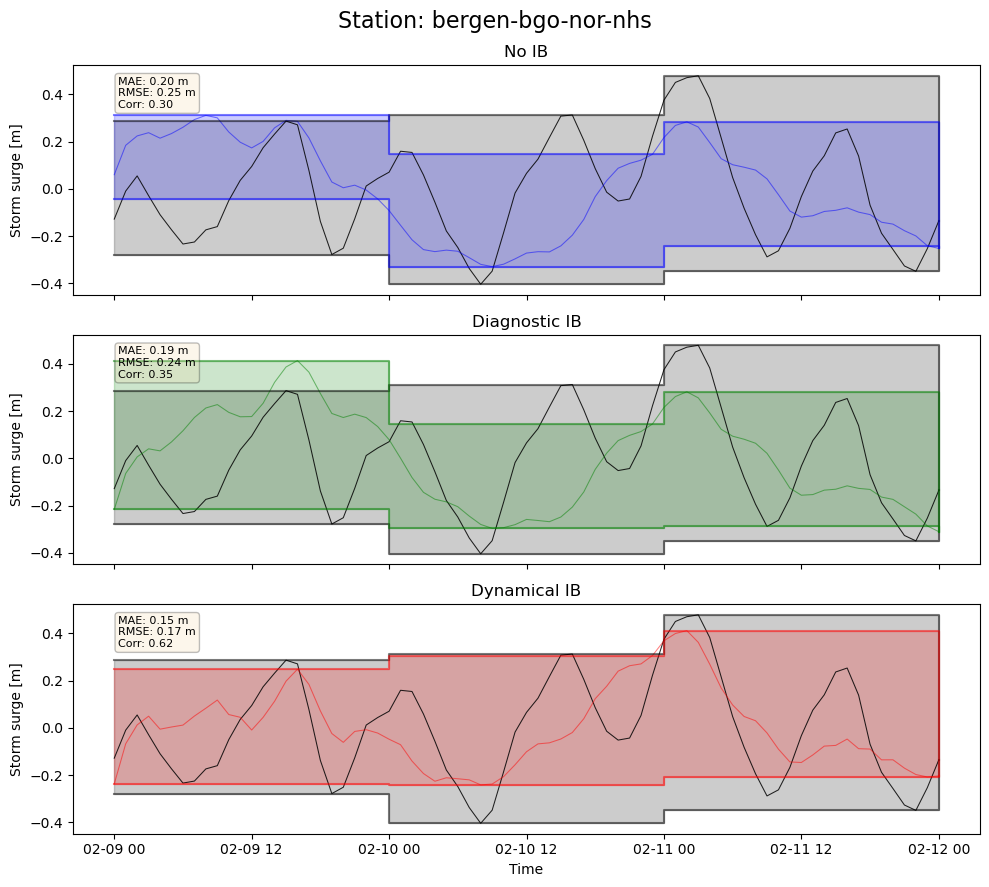}\label{fig:}}  
  \caption{Time series for the stations Cuxhaven (GER) and Bergen (NOR) during the storm Ciara/Elsa (10-11 February 2020). The shaded areas indicate the range between daily minimum and maximum values, while the lines indicate the hourly values for model and observation. The textbox contains the MAE, RMSE and correlation values for the depicted time period for each station an parameter. The vertical reference for the time series is the mean sea level for the shown period.}
  \label{fig:elsa_ciara_timeseries}
\end{figure}

The storm Elsa during 10--11 February 2020 \citep{selberg:etal:2020, jardine2023} was the remnants of a long-lived extratropical cyclone that re-intensified as it passed south of the southern tip of Greenland. The passage eastward followed much the same path as the storm Didrik a month earlier. The maps in Figure \ref{fig:elsa_ciara_map_err} show the average RMSE for the daily maximum storm surge for all stations around the North Sea for the time period 9--12 February. The difference between the \textit{noIB} and \textit{diaIB} experiments are only minor, while the \textit{dynIB} experiment stand out from the two with slightly lower RMSE for the Skagerrak, Norwegian and offshore North Sea stations. The heatmap scatterplot in Figure \ref{fig:elsa_ciara_heatmap_scatter} also point to the \textit{dynIB} as the best performing experiment for the Elsa storm, with both higher correlation and smaller departures from the diagonal. The time series for the four selected stations in Figure \ref{fig:elsa_ciara_timeseries} show how the \textit{dynIB}, most often, does a better job than \textit{noIB} and \textit{diaIB} at simulating both the daily maximum and the hourly water level.

%%%%%%%%%%%%%%%%%%%%%%%%%%%%%%%%%%%%%%%%%%%%%%%%%%%%%%%%%%%%%%%%%%%%%%%%%%%%%%%%%%%%%%%%%%%%%%%%%%%%%%%%%%%%%%%%%%%%%%%%%%%%
\subsection{Tropical cyclones}\label{sec:case:tropical}

\subsubsection{Hurricane Michael}\label{sec:hurricane_michael}
The category 5 Hurricane Michael, on its track from south to north through the Gulf of Mexico, hit the south coast of western Florida near Mexico Beach on the 10 October 2018 \citep{beven2019, kennedy2020}. With winds up to $70\,\mathrm{m\,s^{-1}}$ and a central pressure down to $919\,\mathrm{hPa}$, Hurricane Michael also brought with it an estimated $3\,\mathrm{m}$ storm surge \citep{kennedy2020}.

\begin{figure}
    \centering
    \includegraphics[width=0.75\linewidth]{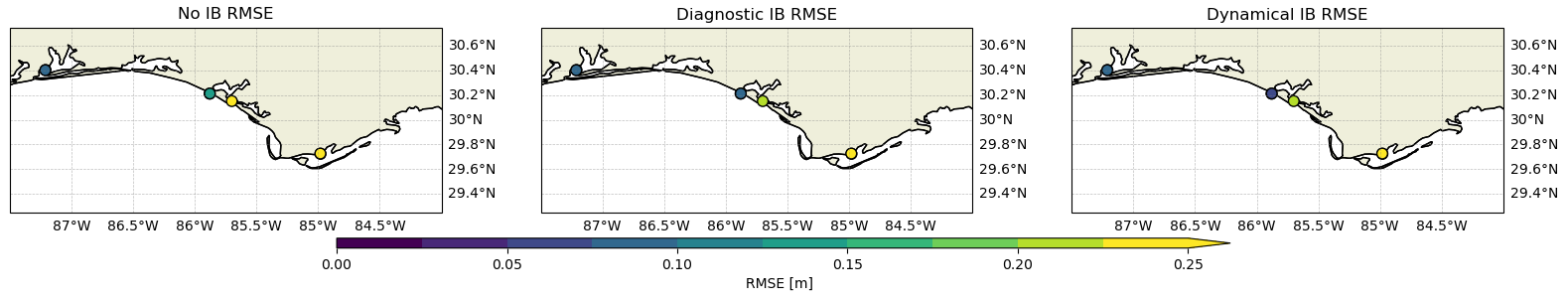}
    \caption{Maps of the Western Florida Coast (US) showing the average RMSE for the daily maximum storm surge during the Hurricane Michael for the time period 7-11 October 2018.}
    \label{fig:michael_map_err}
\end{figure}

\begin{figure}
    \centering
    \includegraphics[width=0.95\linewidth]{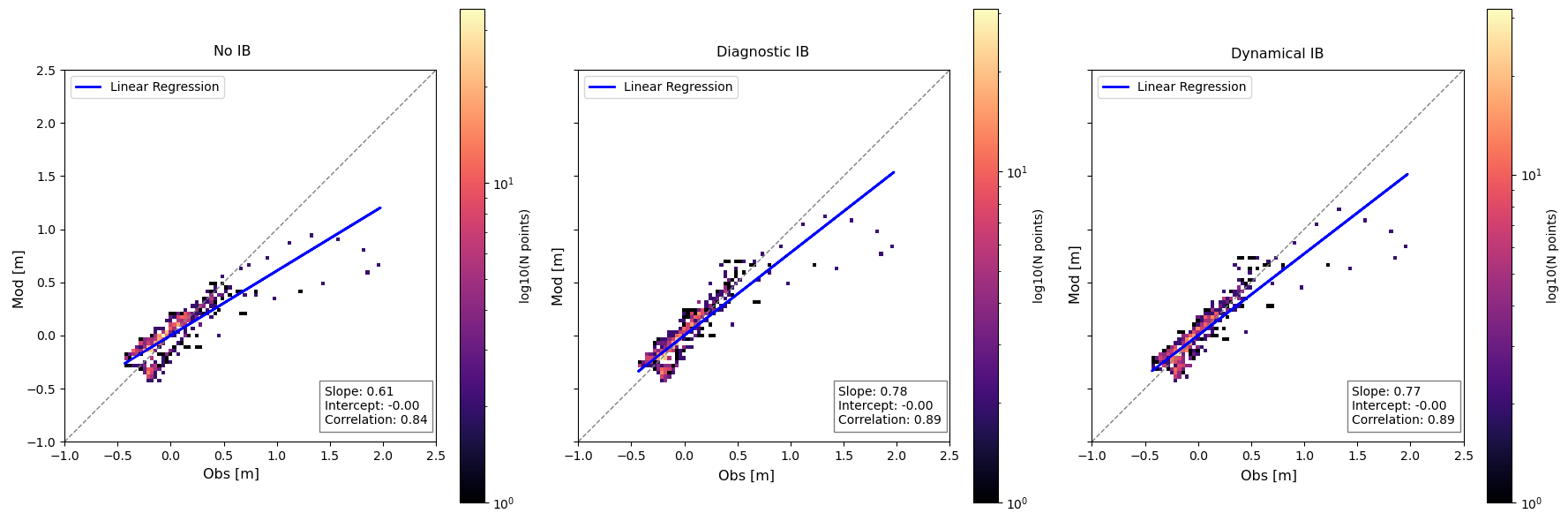}
    \caption{Heatmap scatterplot of the observed and modeled residual water levels for the positions shown in the maps in Figure \ref{fig:michael_map_err} during Hurricane Michael, 7-11 October 2018.}
    \label{fig:michael_heatmap_scatter}
\end{figure}

\begin{figure}[ht]
  \centering
  \subfloat[Apalachicola (FL, USA)]{\includegraphics[width=0.5\textwidth]{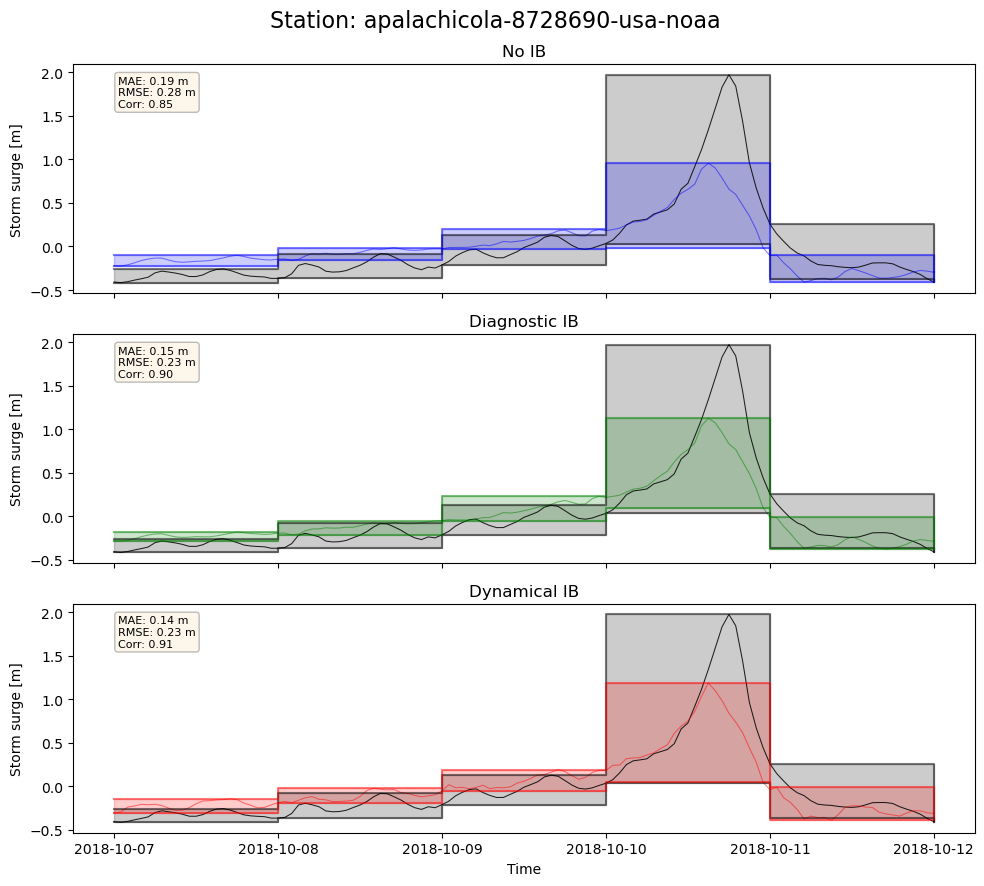}\label{fig:michael_timeseries_a}}
  \subfloat[Panama City Beach (FL, USA)]{\includegraphics[width=0.5\textwidth]{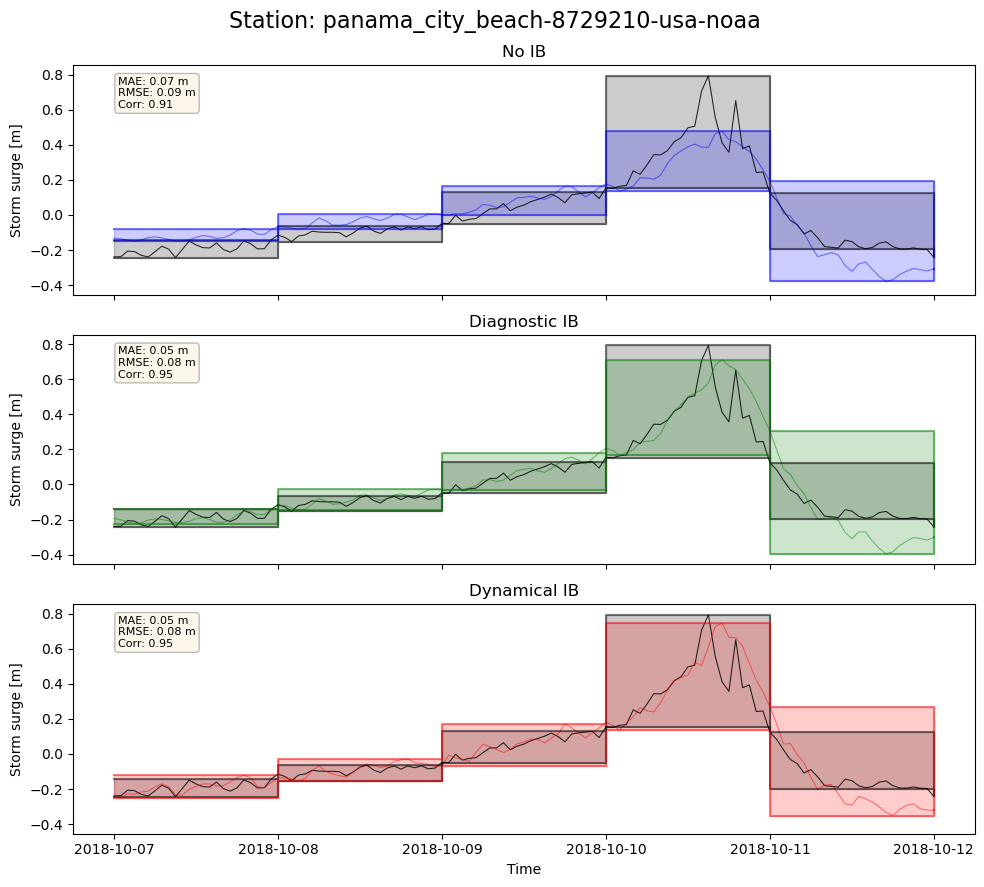}\label{fig:michael_timeseries_c}}
  \caption{Time series for stations Apalachiola and Panama City Beach on the coast of Western Florida during Hurricane Michael. The shaded areas indicate the range between daily minimum and maximum values, while the lines indicate the hourly values for model and observation. The textbox contains the MAE, RMSE and correlation values for the depicted time period for each station an parameter. The vertical reference for the time series is the mean sea level for the shown period.}
  \label{fig:michael_timeseries}
\end{figure}

Figure \ref{fig:michael_map_err} shows the average RMSE for the daily maximum storm surge for the available records along the western Florida coast for our three water level parameters. It is evident how important the IB effect is, but there are no distinguishable differences between the \textit{diaIB} and \textit{dynIB}. 
The same is evident in the heatmap scatterplots in Figure \ref{fig:michael_heatmap_scatter}, where the majority of the data points fall very close to the diagonal for the \textit{diaIB} and \textit{dynIB}, while the \textit{noIB} has slightly larger departures from the diagonal. This is also reflected in the linear correlations, $0.89$ for the \textit{diaIB} and \textit{dynIB} and $0.84$ for \textit{noIB}. The highest observed residual water level in our dataset is approximately $2.0\,\mathrm{m}$ above mean sea level (for the chosen time period), while the model only has a water level of approximately $1.25\,\mathrm{m}$, as seen in Figure \ref{fig:michael_heatmap_scatter} and the time series in Figure \ref{fig:michael_timeseries_a}. However, all the four selected stations in Figure \ref{fig:michael_timeseries} show a marked improvement of both the hourly, and the daily maximum storm surge, with the introduction of the IB effect, but with no clearly visible difference between  \textit{diaIB} and \textit{dynIB}.
We note that even if there is a definite improvement in the results and reduction in error by introducing the IB effect, the highest maximum water level is still significantly underestimated. At the Apalachicola station (see Figure \ref{fig:michael_timeseries_a}, the maximum water level in the model is approximately 1 meter below the observed maximum water level. This is likely caused by the known fact that the ERA5 dataset exhibit generally too low wind speed and not low enough core pressure for tropical cyclones \citep{dulac2024}. Similar underestimation is likely to be expected if examining other tropical cyclone cases.

\subsubsection{Typhoon Mangkhut}\label{sec:typhoon_mangkhut}
Typhoon Mangkhut \citep{ying2022,yang2019} formed over the western Pacific in the beginning of September 2018, and intensified on its path westward towards the northern Philippines, where it made landfall as a category 5 typhoon on 14 September. Moving further westward, the typhoon weakened to a category 1 typhoon before making landfall in southern China, just south of Hong Kong, on 16 September 2018.

\begin{figure}
    \centering
    \includegraphics[width=0.75\linewidth]{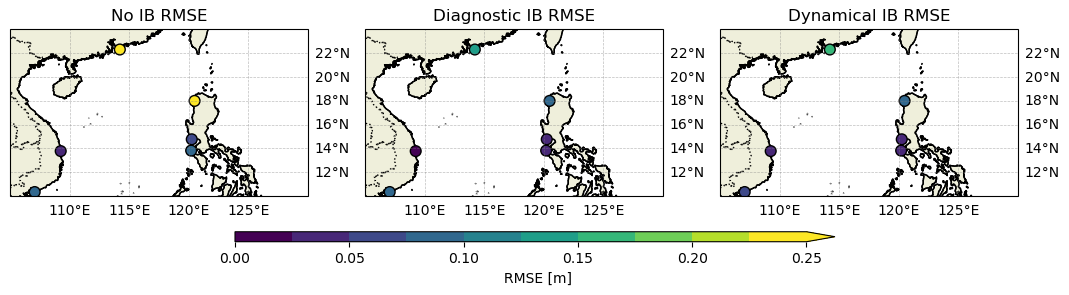}
    \caption{Maps of the South China Sea and adjacent areas showing the average RMSE for the daily maximum storm surge during the Typhoon Mangkhut for the time period 13-17 September 2018.}
    \label{fig:mangkhut_map_err}
\end{figure}

\begin{figure}
    \centering
    \includegraphics[width=0.95\linewidth]{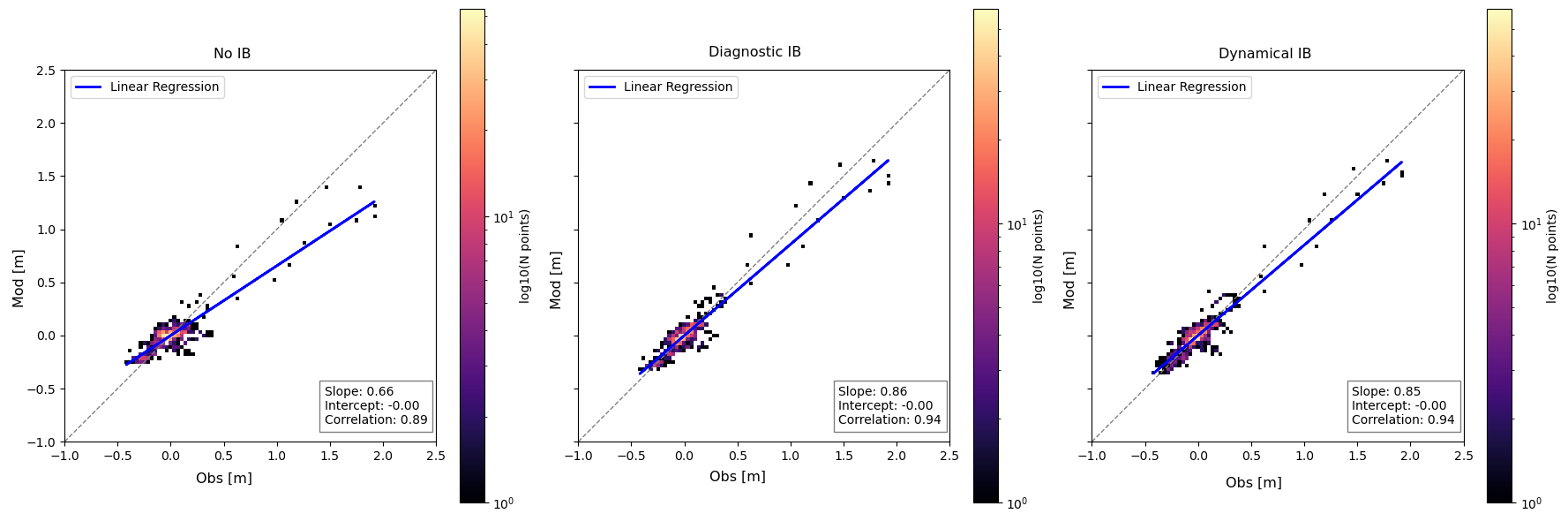}
    \caption{Heatmap scatterplot of the observed and modeled residual water levels for the positions shown in the maps in Figure \ref{fig:mangkhut_map_err} for Typhoon Mankhut, 13-17 September 2018.}
    \label{fig:mangkhut_heatmap_scatter}
\end{figure}

\begin{figure}[ht]
  \centering
  \subfloat[Currimao (PH)]{\includegraphics[width=0.5\textwidth]{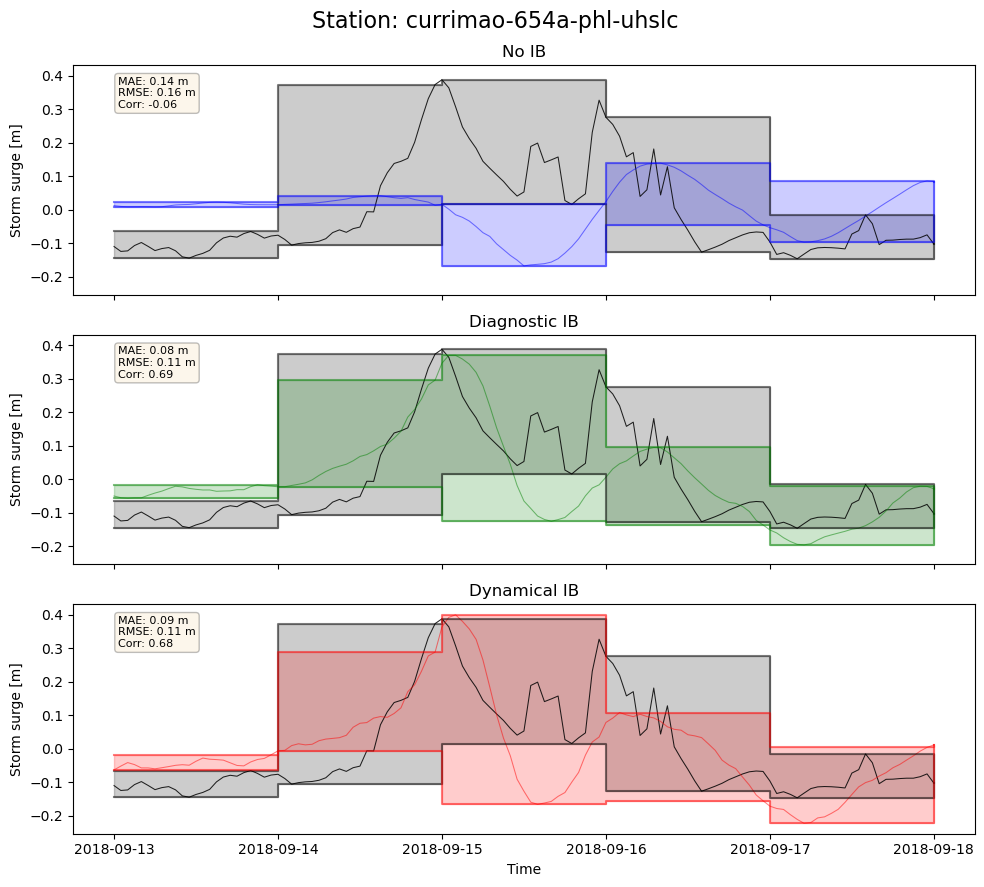}\label{fig:mangkhut_timeseries_a}}
  \subfloat[Hong Kong (CN-HK)]{\includegraphics[width=0.5\textwidth]{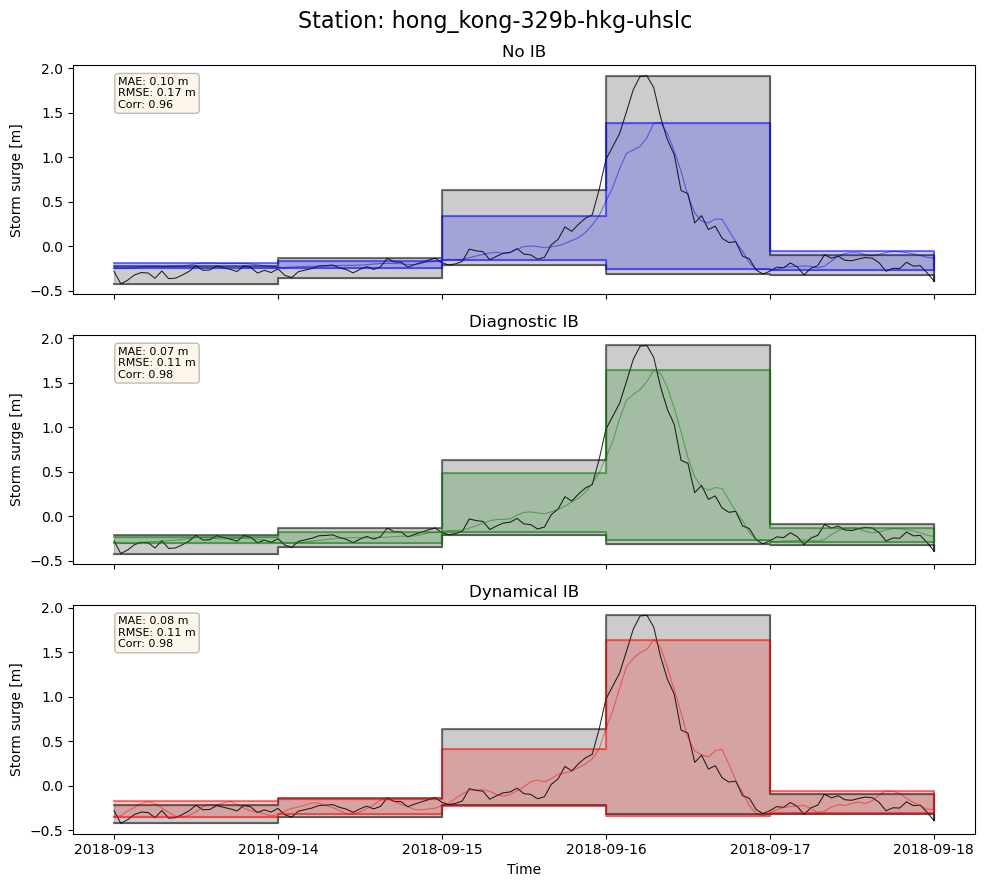}\label{fig:mangkhut_timeseries_b}}  
  \caption{Time series for %two selected 
  stations Currimao (PH) and Hong Kong (CN-HK) around the South China Sea during Typhoon Mangkhut, 13-17 September 2018. The shaded areas indicate the range between daily minimum and maximum values, while the lines indicate the hourly values for model and observation. The textbox contains the MAE, RMSE and correlation values for the depicted time period for each station an parameter. The vertical reference for the time series is the mean sea level for the shown period.}
  \label{fig:mangkhut_timeseries}
\end{figure}

Figure \ref{fig:mangkhut_map_err} shows the average RMSE for the stations located in the South China Sea. The pattern of the errors is similar to what was seen for Hurricane Michael in the previous section, where the introduction of the IB effect clearly reduced the errors. The improvements are also apparent when examining the heatmap scatterplots in Figure \ref{fig:mangkhut_heatmap_scatter}. Further, the time series from the two selected stations in Figure \ref{fig:mangkhut_timeseries} clearly show how the simulations are improved by adding the IB effect. Especially in Figure \ref{fig:mangkhut_timeseries_a}, displaying the time series for Currimao at the north western part of the Philippines, it is obvious how the IB effect is critical to achieve realistic model results.
We note a similar, although not as extreme, underestimation of the highest maximum water level in Figure \ref{fig:mangkhut_timeseries}, as mentioned and explained in the previous Section.

\section{Operational use and implications for storm surge forecasting}\label{sec:ec_ens_use}
We have demonstrated how the implementation of the IB effect in the NEMO model reduces the RMSE for water level in a standalone ocean-only hindcast forced with ERA5 surface winds and atmospheric pressure. Since the NEMO model is the ocean component in the fully-coupled ECMWF earth system forecast model (IFS), it is interesting to explore the impact of including the IB effect in a coupled forecast.  

For operational medium-range forecasts, the IFS is run as an ensemble prediction system \citep[ENS, e.g.][]{ens}, comprising 50 perturbed forecasts and one unperturbed (‘control’) forecast. The perturbed forecasts are designed to represent uncertainties due to the initial conditions and the model integrations (‘model uncertainty’), thus providing a description of forecast probabilities.

The perturbed initial conditions are generated via an ensemble of 4D-Var data assimilations \citep[EDA][]{buizza2008,isaksen2010,lang2019} and singular vectors \citep{leutbecher2008}, which are both added to a central 4D-Var analysis \citep{rabier2000}. In addition, uncertainties in the ocean initial conditions are represented with the use of 11 perturbed ocean analyses \citep{zuo24}.

Model uncertainties are represented via the Stochastically Perturbed Parameterisations scheme \citep[SPP:][]{leutbecher2024, lang2021, ollinaho2017}, which introduces random perturbations to uncertain processes in the atmospheric physics parametrisations throughout the forecast integration. 

To explore the impact of including the IB effect, we ran an ensemble experiment using the operational configuration of the IFS (experiment id io66): one unperturbed and 50 perturbed forecasts, to 15 days. The atmospheric model uses the cubic octahedral reduced gaussian grid TCo1279 (approximately 9km grid-spacing) with 137 vertical levels. The atmospheric model is coupled to the ECMWF Wave Model \citep[e.g.][]{wam}, the \texttt{ORCA025\_Z75} configuration of NEMO V4.0 (as previously described), and the LIM2 sea-ice model \citep{fichefet1997}. 

\subsection{Ensemble simulations of the North Sea storm Xaver}
Storm Xaver moved slowly from west to east across Northern Europe during 5-7 December 2013 \citep{deutschlander:etal:2013, kristensen2024}. The strong northerly winds over the North Sea resulted initially in a convergence of water along the eastern coast of the United Kingdom due to Ekman transport. When the wind direction started to back left, this water started to move anti-clockwise along the North Sea coastline as a free Kelvin wave. The combined Kelvin wave with local amplifications due to Ekman transport by the wind, and the inverse barometer effect resulted in a storm surge among the top five highest recordings over the last 100 years in the German Bight according to \citealt{deutschlander:etal:2013}. 

As the low pressure system moved further east, and the wind direction continued backing further left, the storm surge signal moved further through the North Sea up along the western coast of Denmark and into the Skagerrak area as a Kelvin wave (see \citealt{kristensen2024}). 

\begin{figure}[ht]
  \centering
  \subfloat[Cuxhaven (DE)]{\includegraphics[width=0.75\textwidth]{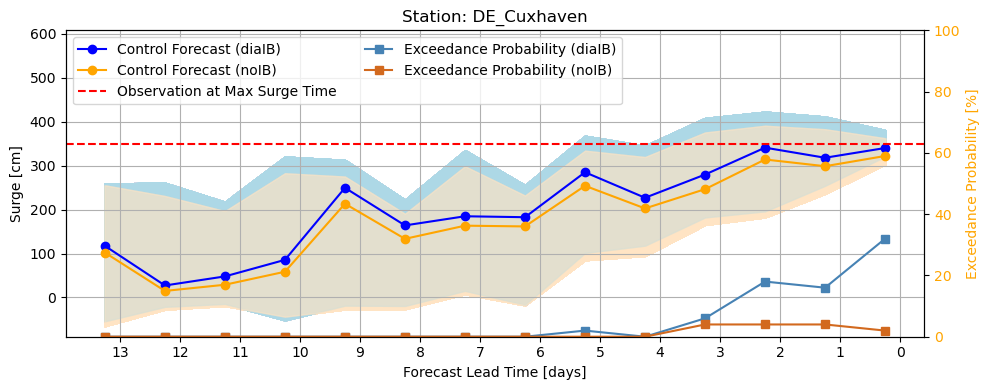}\label{fig:xaver_cux}}
  \hfill
  \subfloat[Hvide Sande (DK)]{\includegraphics[width=0.75\textwidth]{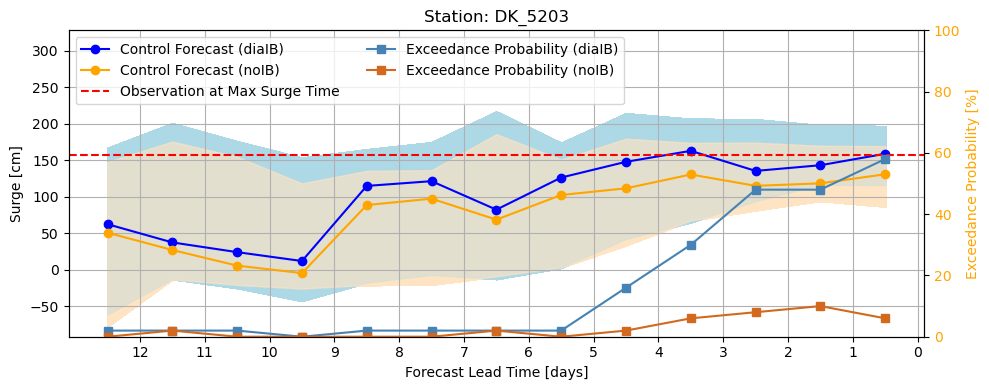}\label{fig:xaver_den}}  
  \caption{Probabilistic ensemble forecasts of residual water levels for the 2013 North Sea Storm Xaver at the peak of the storm surge for two locations. Top panel (a) shows the German station Cuxhaven, located in the German Bight in the southeastern North Sea for maximum surge time (00 UTC), while the bottom panel (b) shows the Danish station Hvide Sande, located at the western coast of Jutland, Denmark, for maximum surge time (06 UTC). The maximum observed residual water level during Storm Xaver for each station is indicated by the dotted red horizontal line. The blue and orange solid lines (with circular markers) show the control member with and without the inclusion of the IB effect, respectively. Similarly, the light blue and peach shaded areas show the ensemble range (from the lowest to the highest values from all ensemble members). The $y$ axis gives the residual water level relative to mean sea level [cm], while the $x$ axis indicates the forecast lead time (in days) prior to the time of the maximum observed residual water level. Please note that since there is only one daily run (starting at 00UTC) of the ENS, and the maximum storm surge occurs at different times at different stations, the forecast lead times will vary between the stations. In addition, the two lowest lines of each plot  (with square markers) indicate the exceedance probability (with the maximum storm surge as criterion) in percent, as expressed by the right side $y$ axis, for the entire ensemble. The steel blue line is with the IB effect taken into account, while the chocolate brown line is without the IB effect.}
  \label{fig:xaver_ens_leadtime}
\end{figure}

The forecasts are initialized daily at 00 UTC from the operational IFS initial conditions for the period from 2013-11-23 to 2013-12-07. To capture the IB effect, the experiment used a research branch based on IFS CY49R2. The IB effect was estimated diagnostically from the mean sea level pressure from the IFS.
Figure \ref{fig:xaver_ens_leadtime} shows how the coupled IFS ENS is able to forecast and give warning about the possibility of this very rare record event up to more than a week in advance. We selected two stations in the area of the highest observed residual water level during this event, namely the German station Cuxhaven, located in the German Bight in the southeastern North Sea, and the Danish station Hvide Sande, located at the western coast of Jutland, Denmark. The ensemble forecast provides an estimate of the uncertainty in a given forecast parameter via the ensemble spread: where the ensemble spread is largest, the uncertainty can be assumed to be greatest (and vice versa). In Figure \ref{fig:xaver_ens_leadtime}, the ensemble spread is indicated by the shaded area that describes the ensemble range --- the difference between the maximum value across the ensemble members and the minimum value. It is evident that by utilizing the entire ensemble, as opposed to only the control member (indicated by the solid lines with the round dots), there is a signal of a maximum surge level as large as the observation more than 5 days in advance at Cuxhaven and throughout the whole forecast period for Hvide Sande.

The exceedance probability, defined as the number of ensemble members, as a percentage, exceeding a certain threshold (in this case, we used the maximum observed residual water level) for a given forecast lead time, is shown as the lower solid lines with square dots in both panels in Figure \ref{fig:xaver_ens_leadtime}. For Cuxhaven, when we include the IB effect, the exceedance probability is larger than zero five days ahead of the event, and increases from three days ahead, to around $35\%$ by day zero. For Hvide Sande, with the IB effect included, the exceedance probability is above zero from more than 12 days ahead of the largest storm surge, with a strong increase from day five to above $50\%$ by day zero. By contrast, the forecasts that do not include the IB effect show exceedance probabilities at both locations of less than $10\%$ even at day zero.

In operational forecasting applications,  information from the real-time ENS runs can be assessed relative to the model climate, which is constructed from hindcasts \citep[see e.g.][Section 5.3]{fug}. This makes it possible to identify cases where the latest ENS forecasts are shifted towards extreme values. The stronger and earlier signal of large maximum storm surge levels from including the IB effect could ultimately provide even earlier warnings of extreme conditions than is indicated from our assessments of absolute forecast values.

%%%%%%%%%%%%%%%%%%%%%%%%%%%%%%%%%%%%%%%%%%%%%%%%%%%%%%%%%%%%%%%%%%%%%%%%%%%%%%%%%%%%%%%%%%%%%%%%%%%%%%%%%%%%%%%%%%%%%%%%%%%%
%%%%%%%%%%%%%%%%%%%%%%%%%%%%%%%%%%%%%%%%%%%%%%%%%%%%%%%%%%%%%%%%%%%%%%%%%%%%%%%%%%%%%%%%%%%%%%%%%%%%%%%%%%%%%%%%%%%%%%%%%%%%
\section{Summary and conclusions}\label{sec:summary}
The purpose of the present work has been to evaluate two methods of including the inverse barometer effect in a stand-alone global ocean model and to assess the potential use of including the IB effect diagnostically in the operational IFS ENS. The model parameter under consideration is the weather-induced water level, and the two methods tested were to either add the IB effect diagnostically as a post process to a model run with constant sea level pressure (\textit{diaIB}), or the direct dynamical inclusion during the model integration (\textit{dynIB}).

The modeled water level was compared against the Global Extreme Sea Level Analysis version 3 (GESLA3), a global dataset of water level observations. The general statistics that include all available model and observation data co-located in time and space yields a MAE of $11\,\mathrm{cm}$ for \textit{noIB}, $10\,\mathrm{cm}$ for \textit{diaIB} and $11\,\mathrm{cm}$ for \textit{dynIB}. The corresponding RMSE values are $14\,\mathrm{cm}$, $13\,\mathrm{cm}$ and $14 \,\mathrm{cm}$, respectively. The weather-induced contribution to the total water level is generally small. To better emphasize how well the three different parameters perform during storm surge events, we masked data where the absolute value of observed residual water level (relative to MSL) was less than one and two standard deviations. This obviously results in increased values for both MAE and RMSE for all three parameters (across all stations considered), with e.g., an RMSE increase of $4\,\mathrm{cm}$ for \textit{noIB}, while both \textit{diaIB} and \textit{dynIB} increase by $2\,\mathrm{cm}$ for the masked data less that one standard deviation. When masking data where observations are less than two standard deviations, the RMSE increases with $11\,\mathrm{cm}$, $5\,\mathrm{cm}$ and $5\,\mathrm{cm}$ for the \textit{noIB}, \textit{diaIB} and \textit{dynIB}, respectively, compared to the average RMSE over all measurements across all stations. 
In addition, the RMSE varies geographically between the stations and between the three parameters, as shown in the histograms in Figure \ref{fig:errdist}. The geographical differences between the \textit{diaIB} and \textit{dynIB} are partly grouped, as seen in Figure \ref{fig:map_rmse_diff_rel} and \ref{fig:map_rmse_diff_rel_eur} where the \textit{diaIB} performs better than the \textit{dynIB} on for example the European North West Shelf, and vice versa in the Baltic Sea. Generally, except for the Baltic Sea, the \textit{diaIB} performs best, or the two perform equally. This is a key finding, since this demonstrates that all ocean models, even those that have no means of including the IB effect dynamically at integration time, can include the IB effect diagnostically as a post process to take into account the pressure effect on the atmospherically induced water level variations.

Our finding that the diagnostic IB addition is often as good as dynamical inclusion globally is consistent with the benchmarks set by \cite{ponte1991} and \cite{wunsch1997}, who found that global departures from isostatic behavior are generally small ($1-3~\mathrm{cm}$). However, the superiority of our \textit{dynIB} experiment in enclosed basins, and during specific storm events, reflects the non-isostatic responses, such as resonance and restricted flow through choking points, documented by \cite{carrere2003} and \cite{gomis2008}.

In addition to the general statistics, a series of case studies were analyzed to highlight the differences between the experiments for a few concrete cases. The time period covered by the model simulation contains several cases of strong extratropical and tropical cyclones. We analyze the two European winter storms Didrik and Elsa (the latter is known as storm Ciara in the UK), together with the cyclones Hurricane Michael and Typhoon Mangkhut. 
For the Didrik and Elsa cases, the combined statistics for all the selected stations indicate that the the \textit{dynIB} is the best performing parameter with both highest correlation, and lowest RMSE. As seen in the maps in Figure \ref{fig:didrik_map_err} and \ref{fig:elsa_ciara_map_err}, this is not the case for all stations, especially for the Skagerrak stations during Didirk, where the \textit{diaIB} and \textit{noIB} has lower RMSE than \textit{dynIB}. Whereas the same stations exert lower RMSE for \textit{dynIB} than \textit{diaIB} and \textit{noIB} during Elsa. Together with the previously mentioned geographical differences, this highlights how one of the alternative methods of including the IB effect is not always better than the other, and despite the fact that the \textit{diaIB} yields the lowest average errors, local considerations should be taken into account.
Based on the analysis of the two tropical cyclones, i.e., Hurricane Michael and Typhoon Mangkhut, there is a key point to be made, namely that although the inverse barometer effect must naturally be included to properly simulate and predict the change in water level due to storm surge, adding it diagnostically as a post process is a perfectly acceptable alternative for coarse models. 

To further assess and exemplify how the inclusion of the IB effect in a global ocean model can improve storm surge early warnings, we show the performance of water level predictions derived from ECMWF ENS forecasts for Storm Xaver. In this case, the model was able to predict the possibility for the very rare and extreme event up to 12 days in advance when including the IB effect.
We note that trying to accurately forecast and predict an extreme event such as Xaver is very challenging for a $0.25^\circ$ global model. Usually, the criterion for issuing warnings of high water level is linked to return periods, such as 5, 50 or 100 years (see e.g. \citealt{kristensen22}). So, the evaluation against the maximum observed storm surge for an event ranked top 5 over the last 100 years, as is done in the present work, should be considered a very strict criterion. 
The Storm Xaver case, and the use of the ensemble prediction system, highlights how the inclusion of the IB effect in the ENS forecasts with the global NEMO model makes the forecasts more capable of providing early warning of extreme storm surge events.

To recapitulate, it is evident that NEMO, a global quarter degree resolution ocean model, forced with ERA5 winds and pressure fields, is capable of reproducing quite faithfully the water level variations associated with the passage of both extratropical and tropical storms, even when compared against water level measurements (detided) in narrow fjords and estuaries. This is encouraging as it suggests that the current setup of ECMWF's IFS can be used as boundary conditions for regional water level forecast systems with the simple addition of the inverse barometer effect. It also shows that an ensemble forecast system such as IFS can be used for early warnings of dangerous storm surges associated with tropical cyclones and extratropical storms.

\printcredits

\section*{Declaration of competing interest}
The authors declare that they have no known competing financial interests or personal relationships that could have appeared to influence the work reported in this paper.

\section*{Declaration of use of generative AI}
During the preparation of this work the authors used Google Gemini in order to research literature during the revision process. After using this service, the authors reviewed and edited the content as needed and takes full responsibility for the content of the published article.

\section*{Open Research Section}
Observations of water level is obtained from the GESLA3 dataset (\url{https://gesla787883612.wordpress.com/}) \citep{haigh2023}.
A dataset containing the extracted time series from the NEMO ocean model runs can be downloaded from \url{https://zenodo.org/records/16894653} \citep{mogensen_2025}. Additionally, the complete model experiments are archived at the ECMWF File System (ECFS) archive as expid i5k6 and i5k8. The ensemble experiment is archived as expid io66.
Code used in the preparation of this manuscript include the python packages pangeo-pytide (\url{https://github.com/CNES/pangeo-pytide}) for performing harmonic analyses, Matplotlib \citep{Matplotlib} for plotting and Cartopy \citep{Cartopy} for creating maps. IPython notebooks for creating all the statistics and plots in this manuscript can be found here: \url{https://github.com/nilsmkMET/nemo_sl_eval}.

\section*{Acknowledgments}
NMK and {\O}B gratefully acknowledge the Research Council of Norway for its support of the StormRisk project (grant no. 300608).

\clearpage

\begin{appendices}
\section{Errors in observations}\label{sec:app:error}
\begin{figure}
  \centering
  \subfloat[Data from BODC]{\includegraphics[width=0.5\textwidth]{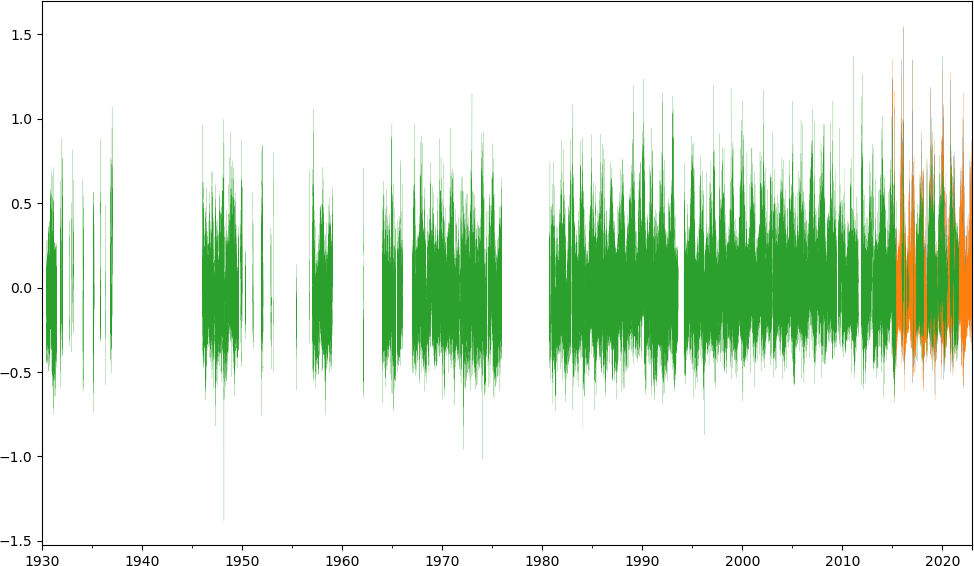}\label{fig:err_ex_a}}
  \subfloat[Data from CMEMS]{\includegraphics[width=0.5\textwidth]{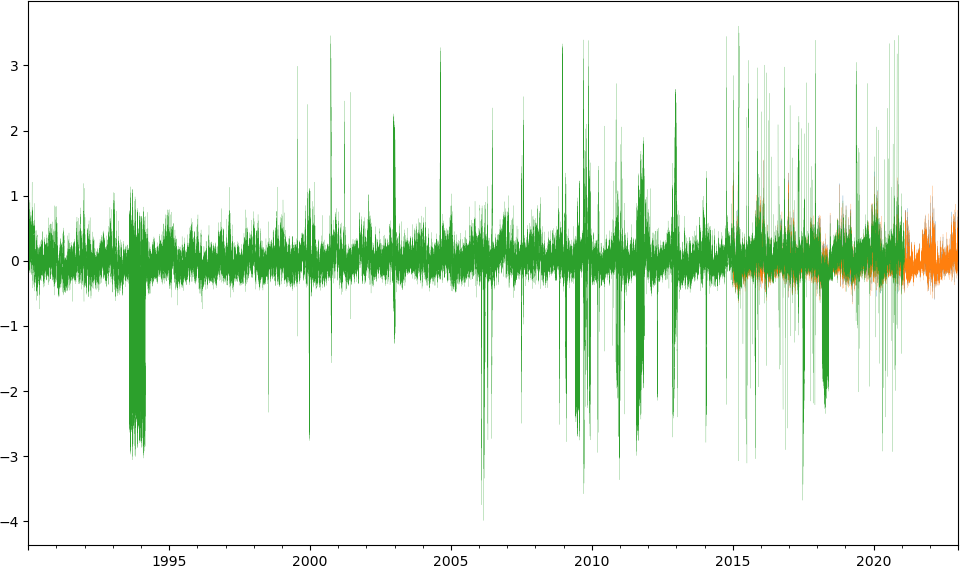}\label{fig:err_ex_b}} 
  \hfill
  \subfloat[Data from BODC]{\includegraphics[width=0.5\textwidth]{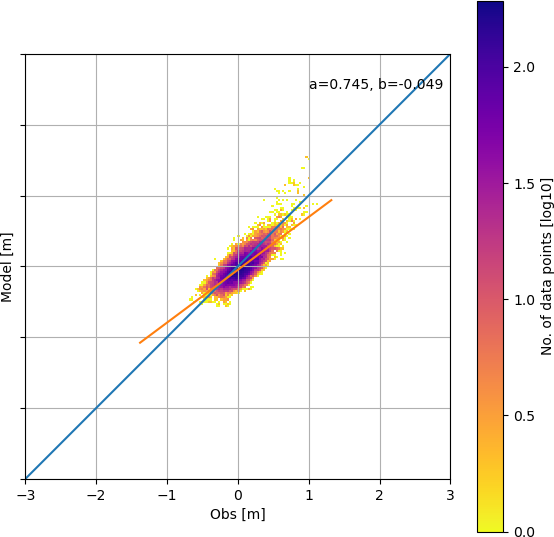}\label{fig:err_ex_c}}
  \subfloat[Data from CMEMS]{\includegraphics[width=0.5\textwidth]{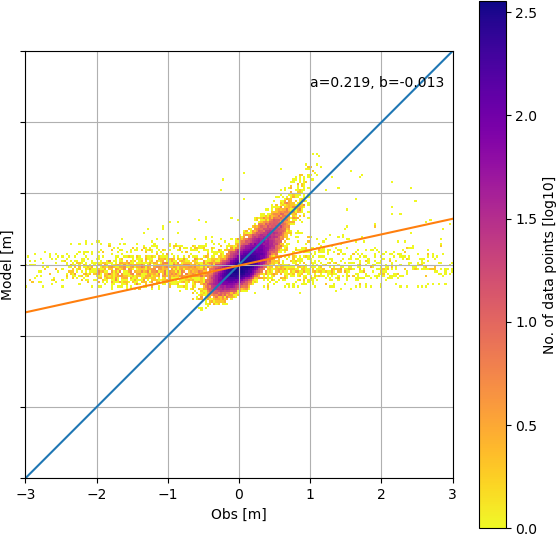}\label{fig:err_ex_d}}  
  \caption{Time interpolation of the observed TWL can create a noisy residual water level signal when subtracting the tides resulting from a harmonic analysis from the TWL. To show this feature we have selected the station of Aberdeen in the United Kingdom. In the GESLA3 dataset this station is represented by a record originating from the British Oceanographic Data Centre (BODC) with a time resolution of combined hourly (pre 1993) and 15 minutes resolution, and  a record from CMEMS with hourly resolution. When inspecting the two, the latter seems to utilize some kind of time interpolation based on the first. The resulting residual water level signal is displayed in the time series in panels (a) and (b) based on the BODC and CMEMS data, respectively. An example of how this influence the validation is shown in panels (c) and (d), where the heatmap scatter plots (on log-scale) show the comparison between the model (Diagnostic IB) and observations based on the BODC and CMEMS data, respectively.}
  \label{fig:error_example}
\end{figure}

As explained in Section \ref{sec:obs_dataset}, there is clearly an effect where resampling of original observation data to, e.g., hourly temporal resolution will introduce small errors in the tidal analysis, and hence result in outliers when subtracting the computed tidal signal from the observed TWL to obtain the residual water level.
An example of this feature is shown in Figure \ref{fig:error_example}, where we show a comparison of the the residual water level signal for the (randomly selected) station Aberdeen (UK) extracted from the observed TWL from the national service British Oceanographic Data Centre (BODC) and the Pan-European CMEMS. 

% \clearpage
\section{Formulas}\label{sec:app:formulas}
The quantities used to assess the error statistics are the mean absolute error (MAE),
\begin{equation}
    \mbox{MAE} = \frac{\sum_{i=1}^{n}\mid \mbox{obs}_i - \mbox{mod}_i \mid}{n},
    \label{eq:mae}
\end{equation}
and the root-mean-square error (RMSE),
\begin{equation}
    \mbox{RMSE} = \sqrt{\frac{\sum_{i=1}^{n} (\mbox{obs}_i - \mbox{mod}_i)^2}{n}},
    \label{eq:rmse}
\end{equation}
as well as their scaled counterparts, $\mbox{MAE}/s$ and $\mbox{RMSE}/s$.
%\begin{equation}
%    \mbox{MAE}/s = \frac{\sum_{i=1}^{n}\mid \frac{\mbox{obs}_i - \mbox{mod}_i}{\sigma} \mid}{n},
%    \label{eq:scaled_mae}
%\end{equation}
%and the scaled RMSE,
%\begin{equation}
%    \mbox{RMSE}/s = \sqrt{\frac{\sum_{i=1}^{n} (\frac{\mbox{obs}_i - \mbox{mod}_i}{s})^2}{n}}.
%    \label{eq:scaled_rmse}
%\end{equation}
Here, the sample standard deviation is computed as
\begin{equation}
    s = \sqrt{\frac{1}{n-1}\sum_{i=1}^{n} (\mbox{obs}_i - \overline{\mathrm{obs}})^2}
    \label{eq:stddev}
\end{equation}
where sample means are indicated by overbars and computed as
\begin{equation}
     \overline{\mathrm{obs}} = \frac{1}{n}\sum_{i=1}^{n} \mbox{obs}_i.
    \label{eq:mean}
\end{equation}

\end{appendices}

\bibliographystyle{cas-model2-names}

\bibliography{surge}

\end{document}